\documentclass[fleqn,usenatbib,onecolumn]{mnras}

\usepackage{newtxtext,amsmath}

\usepackage[T1]{fontenc}

\DeclareRobustCommand{\VAN}[3]{#2}
\let\VANthebibliography\thebibliography
\def\thebibliography{\DeclareRobustCommand{\VAN}[3]{##3}\VANthebibliography}

\usepackage{graphicx}	
\usepackage{amsmath}	
\usepackage{amssymb}	

\newcommand{\soom}{\sim}

\title[Flux Variations on CBPs]{Effects of Flux Variation on the Surface Temperatures of {Earth-analog} Circumbinary Planets}

\author[Yadavalli et al.]{
S. Karthik Yadavalli,$^{1}$\thanks{E-mail: syadavalli6@gatech.edu}
Billy Quarles,$^{1}$ Gongjie Li$^{1}$ and
Nader Haghighipour$^{2}$
\\
$^{1}$Center for Relativistic Astrophysics, School of Physics, 
Georgia Institute of Technology, Atlanta, GA 30332 USA\\
$^{2}$Institute for Astronomy, University of Hawaii-Manoa, Honolulu, HI 96822, USA
}

\date{Accepted XXX. Received YYY; in original form ZZZ}

\pubyear{2020}

\begin{document}
\label{firstpage}
\pagerange{\pageref{firstpage}--\pageref{lastpage}}
\maketitle

\begin{abstract}
The Kepler Space telescope has uncovered around thirteen circumbinary planets (CBPs) that orbit a pair of stars and experience two sources of stellar flux. We characterize the top-of-atmosphere flux and surface temperature evolution in relation to the orbital short-term dynamics between the central binary star and an {Earth-analog} CBP.  We compare the differential evolution of an {Earth-analog} CBP's flux and surface temperature with that of an equivalent single-star (ESS) system to uncover the degree by which the potential habitability of the planet could vary. For a Sun-like primary, we find that the flux variation over a single planetary orbit is greatest when the  dynamical mass ratio is $\sim$0.3 for a G-K spectral binary. Using a latitudinal energy balance model, we show that the ice-albedo feedback plays a substantial role in {(Earth-analog)} CBP habitability due to the interplay between flux redistribution (via obliquity) and changes in the total flux (via binary gyration).  We examine the differential evolution of flux and surface temperature for Earth-like analogs of the habitable zone CBPs (4 Kepler and 1 hypothetical system) and find that these analogs are typically warmer than their ESS counterparts.

\end{abstract}




\section{Introduction} \label{sec:intro}

Observational surveys have uncovered that nearly half of all star systems consist of at least two stars (binary systems) \citep{Raghavan2010,Moe2017}. A planet in a binary star system could either orbit only one star (circumstellar), or around both stars (circumbinary) \citep{Dvorak1982}. Due to the non-constant and non-uniform gravitational potential from the stellar companions, a circumstellar or circumbinary planet (CBP) can experience excitations in the planet's semimajor axis and eccentricity, which can potentially cause a collision with either star or an ejection from the system \citep{Dvorak1986,Rabl1988,Holman1999,Quarles2018,Quarles2020}. Planets that orbit single stars can do so very closely (only a few stellar radii), where a CBP must maintain an orbital distance beyond the critical semimajor axis $a_{crit}$ \citep{Quarles2018} to maintain its orbit over longer timescales. To date, 13 CBPs have been observed: Kepler-16b \citep{Doyle2011Kep16Info}, Kepler-34b and Kepler-35b \citep{Welsh2012Kep35Info}, Kepler-38b \citep{Orosz2012Kep38Info}, Kepler-47b, Kepler-47c \citep{Orosz2012Kep47bc}, Kepler-47d \citep{Orosz2019Kep47dInfo}, Kepler-64b \citep{Schwamb2013Kep64Info}, Kepler-413b \citep{Kostov2014Kep413Info}, Kepler-453b \citep{Welsh2014Kep453Info}, Kepler-1647b \citep{Kostov2016Kep1647Info}, Kepler-1661b \citep{Socia2020Kep1661Info}, and TOI-1338b \citep{Kostov2020TOI1338Info}.

Planetary habitability is parochially defined relative to our own solar system and extended more generally to a range of stellar spectral types in  single-star systems. One guide for habitability is the habitable zone (HZ) that defines the region in which liquid water could exist on the surface of a rocky planet. \cite{KASTING1993} developed an empirical definition for the HZ from observations of Mars and Venus. Venus' surface has been dry for at least 1 billion years \citep{Kasting1983, Solomon1991, KASTING1993,Way2016} and thus a "recent Venus" is used as an empirical inner limit for the HZ. On the other hand, Mars had liquid water 3.8 billion years ago \citep{KASTING1991, KASTING1993,Hurowitz2017}, where "early Mars" denotes an empirical outer limit to the HZ. \cite{KASTING1993} also prescribed a more conservative HZ around a Sun-like star for a planet with an Earth-like $CO_2/H_2O/N_2$ atmosphere and locates the distance at which a runaway greenhouse effect would occur, boiling all liquid water, and the distance at which the maximum possible greenhouse effect would be necessary to keep liquid water from freezing.  \cite{Kopparapu2013} extended their work using updated absorption and scattering coefficients. For our solar system, they found that the empirical HZ begins at 0.75 AU and extends to 1.77 AU, where the conservative HZ has a smaller range of 0.97--1.67 AU.

\begin{table}
\begin{center}
\caption{Coefficients for calculating the spectral weight factor in equation \ref{SpectralWeightAlphaDefined} taken from {\protect\cite{Kopparapu2014}}.}
\label{table:1}
 \begin{tabular}{||c c c c c||} 
 \hline
     & Runaway Greenhouse & Maximum Greenhouse & Recent Venus & Early Mars \\ [0.5ex] 
 \hline\hline
$l_{x-Sun}$ (AU) & {0.95} & 1.67 & 0.75 & 1.77 \\ 
 \hline
 a & {1.332 x 10$^{-4}$} & 6.171 x 10$^{-5}$ & 2.136 x 10$^{-4}$ & 5.547 x 10$^{-5}$ \\
 \hline
 b & {1.580 x 10$^{-8}$} & 1.698 x 10$^{-9}$ & 2.533 x 10$^{-8}$ & 1.5265 x 10$^{-9}$ \\
 \hline
 c & {-8.308 x 10$^{-12}$} & -3.198 x 10$^{-12}$ & -1.332 x 10$^{-11}$ & -2.874 x 10$^{-12}$ \\
 \hline
 d & {-1.931 x 10$^{-15}$} & -5.575 x 10$^{-16}$ & -3.097 x 10$^{-15}$ & -5.011 x 10$^{-16}$ \\ [1ex] 
 \hline
\end{tabular}
\end{center}
\end{table}

Circumbinary habitability has gained interest after the discovery of many  planets through observations from the Kepler mission {\citep{Borucki2010}}. A mechanism has been suggested, where the tidal interaction between binary stars would decrease the rotation of each star, which decreases the total XUV flux and greatly increases a CBP's potential habitability \citep{Mason2013,Mason2015}. \cite{Quarles2012} found that an {Earth-analog} planet could stably orbit in the extended HZ of Kepler-16, where a $CO_2$ dominated atmosphere with increased backwarming would prevent liquid water from freezing. {Following the work of \cite{Kopparapu2013} and \cite{Haghighipour2013}, we define an Earth-analog CBP as a planet with an Earth mass, Earth radius, and a modern Earth-like $CO_2/H_2O/N_2$ atmosphere.} In the case of Kepler-16, one star largely dominates the total brightness of the binary, where \cite{Quarles2012} modeled Kepler-16 as a single-star. \cite{Kane2013} analyzed the habitability of CBPs considering the luminosity and motion of both stars in the binary. They showed that the HZ region shifts over time in response to the periodic motion of the binary stars (i.e., binary gyration). \cite{Haghighipour2013} modeled circumbinary systems considering the weighted flux contribution from each star as a function of the stellar spectral type. Following the methods in \cite{Kopparapu2013}, \cite{Haghighipour2013} accounted for the blackbody peak of each stellar host to compute the fraction of the total output luminosity that will be absorbed by an assumed $CO_2/H_2O/N_2$ atmosphere of the planet, known as the spectral weight factor.  {In our work, we account for the blackbody peak in a similar manner, but use the updated coefficients from \cite{Kopparapu2014} that corrects for N$_2$ partial pressure in their 1D climate model.}  The values for the spectral weight factor ($W$) are uniquely defined for the inner and outer limits of the HZ:
\begin{equation}
    \centering
    W_i^x(T_x,l_x) = [1+\alpha_x(T_i)l_{x-Sun}^2]^{-1}
    \label{SpectralWeightWDefined}
\end{equation}
and
\begin{equation}
    \centering
    \alpha_x(T_i)=a_xT_i + b_xT_i^2 + c_xT_i^3 + d_xT^4_i
    \label{SpectralWeightAlphaDefined}
\end{equation}
where $x$=(in,out) denotes inner and outer HZ limit, $i$=(Pr,Sec) denotes primary and secondary star, \\$l_{x-Sun}$=$(l_{in-Sun},l_{out-Sun})$ denotes the distance from the inner/outer HZ to the Sun, and $T_i(K) = T_{star}(K) - 5780$ represents the difference in effective temperature for each star in Kelvin relative to the Sun. The constants $a_x, b_x, c_x,$ and $d_x$ modulate the weighting factor and are given by Table \ref{table:1}. 

Earth's seasons are due to the daily variation of incident flux on its surface due to the axial tilt, or obliquity. More generally, global flux variations can also arise from a non-zero orbital eccentricity, changes in the distance of a planet from its host star, or periodic modification of the intrinsic luminosity of the host star in addition to the obliquity (i.e., Milankovitch cycles; \cite{Milankovitch1941}).  In the case of high-eccentricity planetary orbits, the flux received by the planet can vary significantly and the planet periodically leaves and re-enters the HZ \citep{Williams2002,Eggl2014,Forgan2015}. In contrast, changes in the surface temperature on a CBP over an orbit depend on flux variations that are influenced by the binary orbit, CBP orbit, and the intrinsic luminosity of each star. The HZ shifts as the location of each star changes in time, and the CBP has a higher probability to venture out of the HZ during its orbit. \cite{Eggl2014} found that even if a CBP's average distance from its host binary is inside the HZ, it could still spend too much time outside the HZ to be habitable as we know it. They argue that to characterize a CBP's habitability, one must account for the dynamics of the system and not only the average received flux. 

{\cite{Forgan2014} showed the range of CBP orbits that lie in the HZ in terms of CBP semimajor axis and eccentricity} for a few circumbinary systems.  \cite{May2016} used both a one-dimensional energy balance model and a three-dimensional general circulation model to evaluate CBP surface temperatures and concluded that circumbinary systems would have very similar surfaces temperatures compared with the equivalent single-star systems. Recently, \cite{Haqq-Misra2019} studied the flux and temperature patterns on a CBP, where they approximate the inner binary using an analytic periodic function. By varying the CBP surface topography, they showed that the CBP surface temperatures can be excited significantly due to the difference in heat capacity between land and oceans.   

In this paper, we further investigate the changes in received flux due to the binary and the resultant surface temperatures on {Earth-analog} CBPs. We find that the flux variations can be largely categorized into two classes depending on orbital and spectral properties of the host binary and CBP. Section \ref{sec:methods} provides a theoretical overview and description of our numerical methods used in our work. The results of our theoretical and numerical models are outlined in Section \ref{sec:results}, with a focus on {Earth-analog} of the known Kepler habitable zone CBPs. A summary along with an overview of possibles avenues of future work are given in Section \ref{sec:summary}. 

\section{Methods} \label{sec:methods}
\subsection{Initial Parameters}
In the interest of comparing to Earth-like habitability, we limit the scope of our investigation to include main-sequence stars with a lifetime of at least a few billion years (i.e., binaries with F, G, K, and M spectral type host stars) in order for a potentially habitable environment and climate to develop.  The dynamical mass ratio $\mu$ $(= M_B/(M_A + M_B))$ of the stellar binary is defined by the masses $M_A$ and $M_B$ of the individual stars, where $M_A \geq M_B$. 

We use the dynamical mass ratio to evaluate general systems, where the luminosity and radius of each star are determined using well-known relationships that correlate with the stellar mass.  The luminosity affects the top-of-atmosphere (TOA) flux a CBP receives from either star, where the stellar radius is necessary to determine the effective stellar temperature through the Stefan-Boltzmann Law ($L/L_\odot = (R/R_\odot)^2 (T_{eff}/T_\odot)^4$).  Table \ref{table:star_params} shows these relations in terms of mass and luminosity. 

\begin{table}
\caption{Stellar luminosity, radius, and effective temperature for a given mass range ({left}) and stellar mass for a given luminosity range ({right})}
\label{table:star_params}
\begin{center}
\begin{minipage}{0.5\linewidth}
 \begin{tabular}{||c c c c||} 
 \hline
  Mass Range   & Luminosity & Radius & $T_{eff}$  \\ [0.5ex] 
   ($M_\odot$) & ($L_\odot$) & ($R_\odot$) & ($T_\odot$) \\ 
 \hline\hline
 $M \leq 0.43$ & 0.23$M^{2.3}$ & $M^{0.54}$ & $0.69 M^{0.305}$ \\ 
  [1ex] 
 \hline
 $0.43<M \leq 1$ & $M^{4}$ & $M^{0.54}$ & $M^{0.73}$ \\ 
  [1ex] 
 \hline
 $1<M \leq 2$ & $1.4M^{3.5}$ & $M^{0.8}$ & $M^{0.6}$ \\ 
  [1ex] 
 \hline
 $M \geq 2$ & 1.40$M^{3.5}$ & $M^{0.8}$ & $1.09 M^{0.475}$ \\ 
  [1ex] 
 \hline
\end{tabular}
\end{minipage}
\begin{minipage}{0.4\linewidth}
\begin{tabular}{||c c||} 
     \hline
      Luminosity Range   & Mass \\ [0.5ex] 
       ($L_\odot$) & ($M_\odot$)\\ 
     \hline\hline
     $L \leq 0.033$ & $(\frac{L}{0.23})^{0.435}$\\ 
      [1ex] 
     \hline
    $0.033<L \leq 16$ & $L^{0.25}$\\ 
      [1ex] 
     \hline
     $16<L \leq 1.72*10^6$ & $(\frac{L}{1.4})^{0.286}$\\ 
      [1ex] 
     \hline
     $L \geq 1.72*10^6$ & $\frac{L}{32000}$\\ 
      [1ex] 
     \hline
    \end{tabular}
    \end{minipage}
\end{center}
\end{table}

\subsection{Flux Calculation}
\label{FluxCalcSection}
\subsubsection{CBP Day Relative to the Center of Mass}
We compute the TOA flux on a planet averaged over one spin period, where we assume the CBP spin period is much shorter than the CBP orbital period (short day). For the Earth, the spin period is the amount of time between successive crossings over the same position in the sky (e.g., zenith) relative to a distant star (sidereal day) or the Sun (solar day).  These periods are not equal because the Earth and Sun are moving relative to the center of mass. A similar distinction between solar and sidereal day can be made for a CBP. However, a CBP's solar day is more accurately called a center of mass (COM) day because the CBP orbits COM of the binary. 
In Figure \ref{fig:COMDayDrawing}a, we sketch the orbit of a CBP relative to the COM of the system. We use the prime meridian vector, which points from the center of the planet through its prime meridian to the binary COM. One COM day is the time interval for one rotation of the planet so that the prime meridian vector returns to pointing toward the COM. Two successive crossings of the primer meridian vector with the COM are denoted by $t_0$ and $t_1$, where the angle $\eta$ is an offset between $t_0$ and $t_1$ that accounts for the spin and orbital motion of the planet. For completeness, one sidereal day corresponds to when these vectors point to a distant background star ($\eta << 1^\circ$), which is a slightly different definition of a day. For the rest of this paper, one "day" is taken to be equal to one COM day, whose duration is set to one Earth day.

\subsubsection{Daily Flux and Spectral Weight Factors}
The instantaneous TOA flux from one star on a short-day planet at a specific latitude is calculated by the following \citep{Armstrong2014,Kane2017}:
\begin{equation}
\centering
F=\frac{L_{*}}{4 \pi d^2} \left(\sin \delta \sin \beta + \cos \delta \cos \beta \cos h \right),
\label{FluxEquation}
\end{equation}
where $L_*$ is the luminosity of the star, $d$ denotes the distance from the star to the planet, $\delta$ represents the stellar declination (i.e., the angle that the star makes with the zenith at the planet's equator), $\beta$ is the planetary latitude, and $h$ is the hour angle (i.e., the angle the star makes with the local meridian in the sky of the planet at a given time).  The leading term in equation \ref{FluxEquation} represents the total flux received by a planet and depends on the distance to each star (i.e., $d_A$ or $d_B$ in Figure \ref{fig:CBPConfig}), where the angular term relates to the altitude of each star above the local horizon for an observer Q (Figure \ref{fig:CBPAngles}). For a CBP, we use an effective luminosity that incorporates the spectral weight factor of each star, using equations \ref{SpectralWeightWDefined} \& \ref{SpectralWeightAlphaDefined} (e.g., $L^\prime_A = W_{Pr}^xL_A$ and $L^\prime_B = W_{Sec}^xL_B$). Then, we calculate the latitudinal flux received by the CBP, from equation \ref{FluxEquation} using 181 latitudinal zones and the weighted luminosity. 

Unless otherwise specified, we assume the planet is near the Runaway Greenhouse limit and use the values in the first column of Table \ref{table:1}. In order to determine the distance $d$ to each star, we model the {orbital dynamics of the three body system} using the \texttt{REBOUND} package with the adaptive-step IAS15 algorithm \citep{Rein2012,Rein2015} {using an initial timestep of 1\% of the binary period.  Since the spin period is one day}, we average equation \ref{FluxEquation} over the range of $[0,2\pi]$ for $h$, the hour angle of each star to determine the daily averaged flux {using the results from the N-body simulation}. The weighted TOA flux contribution received by the CBP from each star ($F_A$ and $F_B$) is calculated using daily averaged form of equation \ref{FluxEquation} and the total TOA flux is the sum of both fluxes (i.e., $F_{tot}(\beta) = F_A(\beta) + F_B(\beta)$) at each latitude.
To characterize the flux variation on the CBP over an arbitrary integration period, at a specific latitude, we evaluate
\begin{equation}
\centering
F_V(\beta)=\frac{F_{max}(\beta)-F_{min}(\beta)}{F_{avg}(\beta)},
\label{FluxVarDefined}
\end{equation}
where $F_{max}(\beta)$ and $F_{min}(\beta)$ are the maximum and minimum TOA flux experienced at the given latitude $\beta$, respectively, and $F_{avg}(\beta)$ is the average of the TOA flux experienced at the same latitude during the integration period.

Flux variations can occur globally for a CBP due to the short period motion of the inner binary.  When $\mu$ is very small, the secondary stellar mass and luminosity are negligible (even when accounting for spectral weights), and the binary approximates a single-star system, which has a relatively small $F_V$. As $\mu$ is increased, the secondary's mass becomes more dynamically important, where the primary's orbit becomes more significantly displaced from the COM (i.e., a larger moment arm relative to the COM).  As a result, the $F_{max}-F_{min}$ from the primary increases from the more significant difference between the maximum and minimum distances from the CBP to the primary star. A star's luminosity is very sensitive to its mass (see Table \ref{table:1}), and the secondary luminosity may be much smaller than the primary luminosity even if the secondary mass is only slightly smaller than the primary mass.  Thus, there is a region of phase space where the secondary star is dynamically relevant but whose luminosity is not.  In this region, the flux variation $F_V$ is larger because the flux amplitude $(F_{max}-F_{min})$ increases faster than the average flux $F_{avg}$.  Once the secondary star's weighted luminosity becomes significant, the average flux $F_{avg}$ increases and the flux variation $F_V$ decreases.  Thus, one expects a peak in $F_V$ at some $\mu$ between 0 and 0.5 (i.e., a curve that is concave down). In Section \ref{FluxVariationsSection}, we show this trend in both an analytically derived approximation and through N-body integrations.  Table \ref{FVOverMuEbinTable} provides coarse values of $F_V$ as a function of the binary orbital parameters ($\mu$ and $e_{bin}$), assuming a Sun-like primary, and $a_{CBP}/a_{bin} = 4$

\begin{table}
\caption{Analytical values of the flux variation $F_V$ as a function of the dynamical mass ratio $\mu$ and binary eccentricity $e_{bin}$.  The binary is composed of a Sun-like primary ($M_A = M_\odot$), where the binary and CBP semimajor axis is 0.1 AU and 0.4 AU, respectively.}
\label{FVOverMuEbinTable}
\begin{center} 
\begin{tabular}{||c c c c c c ||}
\hline
 & $\mu = 0.1$ & $\mu = 0.2$ & $\mu = 0.3$ & $\mu = 0.4$ &$\mu = 0.5$\\ [0.5ex] 
\hline\hline
$e_{bin} = 0.0$ & 0.099 & 0.192 & 0.272 & 0.24 & 0.064 \\ 
\hline
$e_{bin} = 0.1$ & 0.109 & 0.212 & 0.304 & 0.282 & 0.075 \\ 
\hline
$e_{bin} = 0.2$ & 0.119 & 0.233 & 0.336 & 0.324 & 0.086 \\ 
\hline
$e_{bin} = 0.3$ & 0.129 & 0.254 & 0.369 & 0.366 & 0.098 \\ 
\hline
$e_{bin} = 0.4$ & 0.139 & 0.275 & 0.402 & 0.409 & 0.111 \\ 
\hline
$e_{bin} = 0.5$ & 0.149 & 0.296 & 0.435 & 0.452 & 0.125 \\ 
\hline
\hline
\end{tabular}
\end{center}
\end{table}
 
\subsubsection{Stellar Declination}
\begin{figure}
    \centering
    \includegraphics[width=\linewidth]{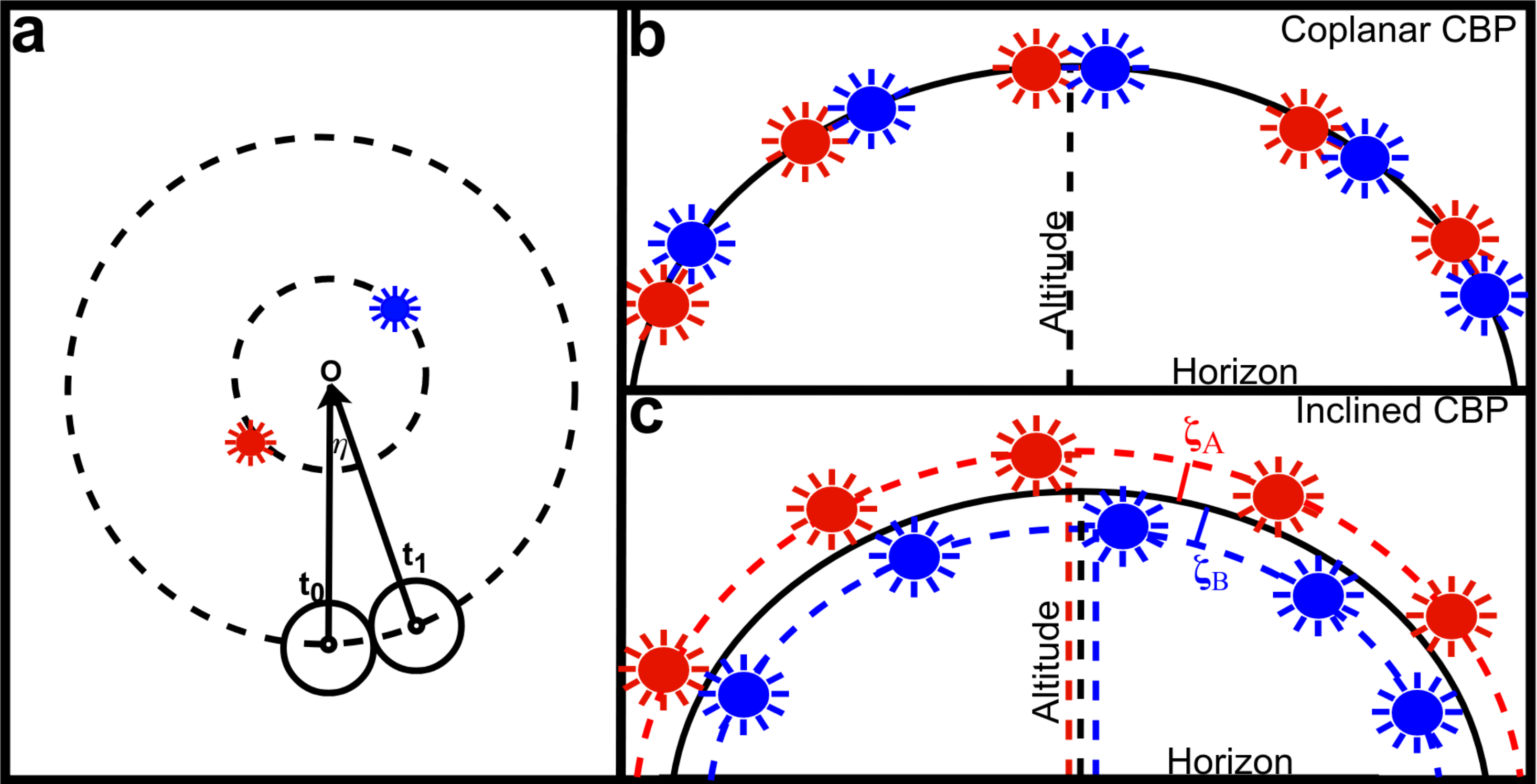}

\caption{A top-down visualization, in panel (a), for a CBP orbiting the center of mass, o, of its primary (red) and secondary (blue) stars, where the definition of a day is relative to the center of mass.  Panel (b) illustrates that the path of the inner binary across the CBP's sky matches the path of the center of mass for a coplanar configuration, where panel (c) shows the differences in altitude for each star when the planetary orbit is inclined relative to the binary.  As a consequence, the integrated flux over a day for an inclined CBP differs from the coplanar configuration by up to $\sim$20 W/m$^2$, where the maximum difference occurs when the CBP and binary orbits are perpendicular to each other.}
\label{fig:COMDayDrawing} 
\end{figure}

In equation \ref{FluxEquation}, $L_{*}$ is dependent on the host stars and $d$ is obtained through the relative position vector between the planet and each star. The declination $\delta$ for a single star system \citep{Armstrong2014} is given by 
\begin{equation}
    \delta = \psi  \sin L_s,
    \label{declSingleStar}
\end{equation}

and 

\begin{equation}
    L_s = L_{sp} + f
    \label{solarLongPeri}
\end{equation}
where $\psi$, $L_{sp}$, and $f$ are the planetary obliquity, stellar longitude of perihelion, and CBP true anomaly, respectively. In effect, equation \ref{declSingleStar} finds the declination of the system's COM. If the CBP's orbit is coplanar with the central binary (no mutual inclination), then each star takes the same path across the CBP sky during one day as the COM path, where the declination of both stars is equal to the declination of the COM (see Fig. \ref{fig:COMDayDrawing}b). In the case of an inclined CBP, the path across the sky for each star is different (see Fig \ref{fig:COMDayDrawing}c), making the daily averaged flux for each star different.  We extend equation \ref{declSingleStar} for finding the declination of a star on an inclined CBP with the addition of a term that represents the angular separation $\zeta_X$ of each star from the COM path: 
 
 \begin{equation}
     \delta_X = \left(\psi + \zeta_X\right) \sin L_s,
     \label{declCBP}
 \end{equation}
where $\zeta_X$ is determined using the projected $z$-component for each star $z_X$, the line connecting the COM to the CBP's equator as the reference plane, and the distance $r_{CBP}$ from the COM to the CBP ($\zeta_X = \arctan[ z_X/r_{CBP}]$). The addition of this extra term accounts for any differences in stellar declination due to the CBP's mutual inclination relative to the COM declination. We test the extent of this correction numerically and find a maximum difference of $\soom$ 5-20 W/m$^2$ in the TOA flux from a CBP inclined by $10^\circ-90^\circ$, respectively, compared to the coplanar CBPs. The differences in daily flux due to this correction are small and the known CBPs are nearly coplanar with their host binaries. Thus, we consider coplanar CBPs in this work and use the declination given by equation \ref{declSingleStar} as the declination for each star.

\subsubsection{CBP Flux Categories: \textit{Case 1} and \textit{Case 2}}
The TOA flux received by any planet from its star varies with distance $d$ between planet and star as $F \soom 1/d^2$. A planet orbiting a single star therefore necessarily experiences $F_{max}$ during pericenter passage and $F_{min}$ during apocenter passage. If the orbit is not very eccentric, $F_{max}$ and $F_{min}$ are close to each other and $F_V$ is very small. In contrast, a CBP does not necessarily experience $F_{min}$ during its apocenter passage or $F_{max}$ during pericenter passage due to the changing relative alignment between the binary host stars and the CBP, . The incident flux on the CBP depends on the effective temperature of each star, the relative orientation of the CBP and binary orbits, and the true anomalies of the CBP and binary. Even if the CBP's orbit has very little eccentricity, the distance to each star can vary considerably, meaning that the difference between $F_{max}$ and $F_{min}$ can still be quite large. Thus, for a given planet eccentricity, $F_V$ is larger for a CBP than for a single-star planet.

Two cases arise when considering the variation in a CBP's TOA flux that differ in timescale. In \textit{Case 1}, the secondary star's luminosity is comparable to the primary star's luminosity ($L_B \soom L_A$).  There are two maxima for the CBP incident flux, where the first occurs when the primary star eclipses the secondary star and a secondary peak occurs when the secondary star eclipses the primary. The minimum CBP incident flux occurs at a point in between the maxima, when neither star is fully eclipsed and the CBP is closer to the secondary than to the primary. For this case, the CBP flux varies with respect to half the binary timescale ($P_{bin}/2$) and the CBP orbital period ($P_{CBP}$). 

In \textit{Case 2}, the secondary star's luminosity is much smaller than the primary star's luminosity ($L_B << L_A$).  The maximum incident flux occurs when the CBP is closest to the primary and the minimum incident flux occurs when the CBP is farthest from the primary star, minimizing the secondary star's contribution. For \textit{Case 2}, the TOA flux varies on the binary timescale ($P_{bin}$) and the CBP timescale ($P_{CBP}$). In both cases, the CBP's forced eccentricity creates a $P_{CBP}$-periodicity in CBP flux. \cite{Popp2017} present a similar analysis of a hypothetical habitable planet in the Kepler-35 system (\textit{Case 2} in our scheme), where they show that such a planet would receive maximum flux when the primary eclipses the secondary and minimum flux when the secondary eclipses the primary.  Figure \ref{fig:FluxMaxMinCBP} highlights these configurations, showing the primary star in red and secondary in blue.

\subsubsection{Border Between Case 1 and Case 2} \label{MathInterlude}
In this section, we derive an approximate expression for the border between \textit{Case 1} and \textit{Case 2}. Figure \ref{fig:FluxMaxMinCBP}a illustrates a \textit{Case 1} configuration with a primary maximum (Fig. \ref{fig:FluxMaxMinCBP}a-I), where the secondary luminosity contributes to the secondary maximum in flux (Fig. \ref{fig:FluxMaxMinCBP}a-III) and a minimum occurs between the two maxima (Fig. \ref{fig:FluxMaxMinCBP}a-II). As the secondary luminosity decreases, the secondary maximum also decreases and transforms into a minimum (Fig. \ref{fig:FluxMaxMinCBP}b-V), while the previous minimum (Fig. \ref{fig:FluxMaxMinCBP}a-II) becomes an inflection point (Fig. \ref{fig:FluxMaxMinCBP}b-IV). There is a border between \textit{Case 1} and \textit{Case 2}, where the flux received at Fig. \ref{fig:FluxMaxMinCBP}a-II and Fig. \ref{fig:FluxMaxMinCBP}a-III is equal, and

\begin{equation}
\left. \frac{\partial F}{\partial \theta} \right\rvert_{\theta = \pi} = \left. \frac{\partial^2 F}{\partial \theta^2} \right\rvert_{\theta = \pi} = 0,
\label{equ:derivativesFluxBorder}
\end{equation}

where $\theta$ is the angle between $r_a$, the position vector of the primary star, and $r_{CBP}$, the position vector of the CBP (see Figure \ref{fig:CBPConfig}). Writing the flux in terms of $\mu$, $M_A$, $a_{bin}$, $a_{CBP}$, $e_{bin}$, and $e_{CBP}$ and by solving equation \ref{equ:derivativesFluxBorder} (see Appendix \ref{BorderDeriveAppendix}), we find the following for the luminosity ratio:

\begin{equation}
\frac{L_B}{L_A} = \frac{\mu [1 + \alpha (e_{bin} - 4) (e_{bin} + 1) \mu]}{[1 - \alpha (e_{bin} - 4) (e_{bin} + 1) (1 - \mu)] (1 - \mu)}.
\label{eq:BorderEquation}
\end{equation}

Equation \ref{eq:BorderEquation} identifies the ratio $L_B/L_A$ for a border case in terms of the orbital parameters of the circumbinary configuration. For main-sequence stars, one can prescribe a primary mass and use the mass-luminosity relation to numerically solve for the value of the dynamical mass ratio ($\mu_o$) at the border between \textit{Case 1} and \textit{Case 2}. Then, any $\mu < \mu_o$ is categorized as \textit{Case 2} and any $\mu > \mu_0$ is \textit{Case 1}. For the remainder of the paper, we use this method to determine $\mu_o$ when discussing the border between cases.

\section{Results} \label{sec:results}

\subsection{CBP Flux Patterns}
We investigate the TOA flux received by a CBP and how it is influenced by the system's dynamical parameters. Figure \ref{fig:Earth-CBPDiff} shows the daily averaged TOA flux (color coded) received by an Earth-like planet ($F_{tot}$ = 1361 W/m$^2$, or 1 S$_\oplus$) orbiting the Sun (Fig. \ref{fig:Earth-CBPDiff}a), a \textit{Case 1} CBP ($\mu = 0.5$; Fig. \ref{fig:Earth-CBPDiff}b), and a \textit{Case 2} CBP ($\mu = 0.09$; Fig. \ref{fig:Earth-CBPDiff}c). Figures \ref{fig:Earth-CBPDiff}b and \ref{fig:Earth-CBPDiff}c have bright and dark fringes due to periodic changes in flux on two timescales (see Section \ref{FluxCalcSection}). As expected, the frequency of the fringes is higher for \textit{Case 1} than \textit{Case 2}.

As a planet's obliquity increases, the daily TOA flux becomes more concentrated at higher latitudes. For very high obliquities, only the poles are exposed to direct light for a large part of the orbit, resulting in dramatic changes over a planetary orbit. Figure \ref{fig:FluxCBPSemimObl} shows the daily averaged TOA flux (color coded) received by a CBP with an Earth-like obliquity (23.5$^\circ$; Fig. \ref{fig:FluxCBPSemimObl}a), 45$^\circ$ (Fig. \ref{fig:FluxCBPSemimObl}b), and 90$^\circ$ (Fig. \ref{fig:FluxCBPSemimObl}c). The daily TOA flux for a CBP resembles a single star in terms of flux changes due to the obliquity, except that fringes appear due to the motion of the inner binary. As the semimajor axis ratio ($a_{CBP}/a_{bin}$) increases, the fringes due to the binary blend together and the resulting daily TOA flux more closely resembles the equivalent single-star configuration.  Figure \ref{fig:FluxCBPSemimObl}d--\ref{fig:FluxCBPSemimObl}f shows the daily averaged TOA flux (color coded) for a CBP with increasing semimajor axis relative to the stability limit ($a_{CBP}/a_{crit}$), with values of 2, 3, and 4, respectively.  There are more binary orbits per single CBP orbit when the CBP semimajor axis is increased, which increases the number of bright fringes and making each bright fringe less pronounced. At large $a_{CBP}/a_{bin}$, the flux for a CBP is indistinguishable from that of the equivalent single-star case.

\subsection{Flux Variations}
\label{FluxVariationsSection}
Kepler-16AB  (see Table \ref{table:Plot8CBPValues}) hosts at least one circumbinary planet, Kepler-16b, which is $\sim0.7$ AU from the binary on a nearly circular orbit \citep{Doyle2011Kep16Info}. The luminosity of Kepler-16B is much less than the luminosity of Kepler-16A ($L_B/L_A \approx 0.01$), making Kepler-16b an example of \textit{Case 2} (see {section} \ref{sec:methods}). We note that the TOA flux is independent of the assumed planetary size or atmospheric composition. In Figure ~\ref{fig:FluxSA250}, we examine the daily TOA flux at the CBP equator during a single orbit. Kepler-16b, a CBP with an almost circular orbit ($e=0.007$) still experiences significant variability in its  incident flux, which supports our claim that even CBPs with small eccentricities can experience significant flux variations. These variations are driven by two phenomena: the change in flux due to the orbit of the binary and a change in flux due to the CBP's forced non-zero eccentricity, where the variation due to only the binary's orbit is 34\% and accounting for the CBP's eccentricity increases the overall variation to 43\%.

The flux variation received by a CBP depends on the inner binary moment arm relative to the COM through the dynamical mass ratio $\mu$, primary mass $M_A$, binary eccentricity $e_{bin}$, and the semimajor axis ratio $a_{CBP}/a_{bin}$. In Figure \ref{fig:FluxVarMu}, we show how the flux variation changes with respect to each of these factors. Figure  \ref{fig:FluxVarMu}a illustrates the simulated value of $F_V$ using an N-body simulation and an analytically computed value of $F_V$, where we use equations \ref{eqn:cosine_law}--\ref{eqn:Leg_exp_sqr} to approximate $F_V$.  The simulation assumes a Sun-like primary star while the secondary star's mass varies across the full range of $\mu$ ($0 < \mu \leq 0.5$), with a mildly elliptical orbit ($e_{bin} = 0.13$), and a typical binary semimajor axis ($a_{bin} = 0.15$) of the known CBP host binaries. The CBP starts in a circular orbit with a semimajor axis ratio ($a_{CBP}/a_{bin} = 3.84$) that is beyond the critical value \citep{Quarles2018}.

There is a peak in the value of $F_V$ at approximately $\mu=0.33$ in Figure \ref{fig:FluxVarMu}a, corresponding to where the secondary star's dynamical influence is most important and luminosity is $\lesssim 0.06$ L$_\odot$  (see Section \ref{MathInterlude}). The differences in the analytically predicted $F_V$ and N-body simulated $F_V$ are the result of planet eccentricity excitations due to mutual gravitational attractions between the binary and CBP. The flux variation $F_V$ also depends slightly on the primary mass $M_A$, where we determine the flux variation for a primary mass of $0.3 M_\odot$, $0.8 M_\odot$, and $1 M_\odot$ that corresponds to three spectral types (M, K, and G, respectively) in Fig \ref{fig:FluxVarMu}b. The peak in $F_V$ shifts to different values of $\mu$, where an increase in the primary mass results in a higher peak value in $\mu$ and vice versa. As the CBP's distance from the binary increases, the binary star appears more like a single star and the flux variation decreases. Figure \ref{fig:FluxVarMu}c shows this expected trend with the semimajor axis ratio ($a_{CBP}/a_{bin}$) using a Sun-like primary, where the secondary mass is determined via the dynamical mass ratio $\mu$ and the curves correspond to $\mu$ values of 0.25 (gold), 0.33 (green), and 0.5 (blue).

A higher binary eccentricity increases the maximum flux and decreases the minimum flux received by the CBP, while holding the average flux roughly constant. This implies that $F_V$ always increases with binary eccentricity. As detailed in section \ref{MathInterlude}, $F_V$ also depends on the luminosity and mass of both stars. Thus, $F_V$ depends on $\mu$ and binary eccentricity independently. As a result, the peak of flux variation $F_V$ in $\mu$ does not shift much with binary eccentricity. Across the full parameter space explored in Figure \ref{fig:FluxVarMu}d, the maximum critical semimajor axis is $3.84 a_{bin}$ \citep{Quarles2018} for a mass ratio $\mu$ of 0.24 and binary eccentricity 0.49. In Figs. \ref{fig:FluxVarMu}a, b, and d, the CBP begins on a circular orbit with a semimajor axis of $3.84 a_{bin} =  0.576 AU$, so that the semimajor axis ratio $a_{CBP}/a_{bin}$ is not varied and stability is guaranteed for all simulations in these three panels. As expected from Fig. \ref{fig:FluxVarMu}c, changing $a_{CBP}/a_{bin}$ changes the magnitude of $F_V$ in Fig. \ref{fig:FluxVarMu}d, but the structure of the plot is conserved. 

The cyan line shows the location of the border between \textit{Case 1} and \textit{Case 2} for a CBP semimajor axis of $3.84 a_{bin}$. A change in the semimajor axis ratio $a_{CBP}/a_{bin}$ changes the position of the line, where a larger $a_{CBP}/a_{bin}$ moves the line to the right and vice versa. If we use the CBP semimajor axis of the actual Kepler systems, we find Kepler-34b and Kepler-35b are the only \textit{Case 1} configurations of the currently known CBPs. Our results suggest that \textit{Case 1} configurations will be much less common than \textit{Case 2} configurations because the fractional area of phase space corresponding to \textit{Case 2} is so much larger. Although there is an excess of twin stars ($\mu \sim 0.5$) with respect to mass ratio shown in binary star surveys, the cumulative fraction of Solar-type primaries with low mass secondary stars is large \citep{Raghavan2010,Moe2017}.

\begin{table}
\caption{Orbital and Spectral Properties of Binary Systems with CBPs in the conservative HZ. Values for Kepler-16b, Kepler-47c, Kepler-453b, and Kepler-1647b are taken from \protect\cite{Doyle2011Kep16Info}, \protect\cite{Orosz2012Kep47bc}, \protect\cite{Welsh2014Kep453Info}, and \protect\cite{Kostov2016Kep1647Info}, respectively. The 'Extreme CBP' is a hypothetical case where $a_{CBP}$ is near the inner edge of stability and its HZ.}
\label{table:Plot8CBPValues}
\begin{center}
 \begin{tabular}{||c c c c c c c c c c c c||} 
 \hline
     & $M_A (M_\odot)$ & $M_B (M_\odot)$ & $L_A (L_\odot)$ & $L_B (L_\odot)$ & $T_A $(K) & $T_B$ (K) & $a_{bin}$ (AU) & $e_{bin}$ & $a_{CBP}$ (AU) & $e_{CBP}$ & $F_V$ \\ [0.5ex] 
 \hline\hline
 Kepler-16b  & 0.69 & 0.2 & 0.148 & 0.0057 & 4446 & 3311 & 0.22 & 0.16 & 0.705 & 0.007 & 0.239\\ 
 \hline
 Kepler-47c & 1.043 & 0.362 & 0.84& 0.014 & 5636 & 3357 &0.0836 & 0.0234 & 0.99 & 0.041 & 0.258\\
 \hline
 Kepler-453b & 0.93 & 0.19 & 0.58 & 0.00474 & 5527 & 3309 & 0.18 & 0.051 & 0.79 & 0.038 & 0.285\\
 \hline
 Kepler-1647b & 1.22 & 0.97 & 4.271 & 0.93 & 6210 & 5770 & 0.13 & 0.16 & 2.72 & 0.058 & 0.291\\
 \hline
 Extreme CBP & 0.62 & 0.305 & 0.1478 & 0.015 & 4077 & 2776 & 0.15 & 0.11 & 0.40 & 0.0 & 0.446\\
 \hline
 \hline
\end{tabular}
\end{center}

\end{table}

\subsection{Energy Balance Models}
We extend our numerical investigations of TOA flux to examine the possible surface temperature variations for {Earth-analog} CBPs compared with the respective equivalent single-star systems. One can evaluate the temperature by solving the following 1D Latitudinal Energy Balance Model (EBM) for CBP surface temperature (T) as a function of time ($t$) and latitude ($\beta$):
\begin{equation}
    C\frac{\partial T}{\partial t} = S(1-\alpha) - (A+BT) + \frac{1}{\cos{\beta}}\frac{\partial}{\partial \beta}(D \cos{\beta}\frac{\partial T}{\partial \beta}),
    \label{equ:EBM}
\end{equation}

where $S$ is the TOA flux received at a specific latitude at each sample (1 Earth day) of an N-body simulation, and is evaluated using the process described in Section \ref{FluxCalcSection}. Equation \ref{equ:EBM} depends on the assumed heat capacity $C$, where we use heat capacity of water. The constants $A$ and $B$ are the thermal emission parameters: $210$ W/m$^2$ and $2$ W/m$^2$/K, respectively, {and} $D$ represents the latitudinal diffusive parameter, $0.55$ W/m$^2$/K. The model assumes a cloudless water-world with a uniform heat capacity at all latitudes and an equivalent water depth of 10 meters.  This is appropriate because none of the known CBPs are expected to be terrestrial and we expect that our work will become more applicable once {Earth-analog} CBPs are discovered.  We also explore the difference in surface temperatures compared with the equivalent single-star configuration for each of our {Earth-analog} CBP analog, where any systematic bias due to our model assumption is mitigated. The albedo $\alpha$ is defined as:
\begin{equation}
    \alpha = \begin{cases} 
      0.6, & T \leq 263 K \\
      0.3 + 0.078 \mathcal{P}_2(\sin \beta), & T > 263 K, \\
   \end{cases}
       \label{equ:albedo}
\end{equation}

where $\mathcal{P}_2$ is the second Legendre polynomial. We use the \texttt{CLIMLAB} package in python as described by \cite{Rose2018} to evaluate the EBM. The EBM decomposes the CBP into 181 latitudinal zones, assumes a surface temperature profile, and then numerically solves the EBM using a timestep of approximately $8.64\times 10^3$s or 1 Earth day. Further details concerning the EBM are thoroughly described in the documentation for \texttt{CLIMLAB} \citep{Rose2018}.

{We used a simple 1D EBM that parameterizes the incoming and outgoing radiation, which is in contrast to the EBM used by \cite{Cukier2019} that evaluates the HZ limits more directly from models of the assumed atmospheric chemistry \citep{Kopparapu2013}.  \citeauthor{Cukier2019} also used the CBP setup prescribed by \cite{Kane2013} to modulate the flux for CBPs in various binary stellar types. Three-dimensional EBMs could be used to evaluate longitudinal energy flow, more detailed atmospheric dynamics, and provide more realistic results, but this is beyond the scope of our work.}

\subsubsection{Case 1 CBPs} \label{sec:case1_EBM}

\begin{table}
\caption{Values for \textit{Case 1} and \textit{Case 2} to evaluate our energy balance model (EBM).  The binary spectral parameters ($M$, $L$, and $T$) are chosen so that an {Earth-analog} CBP receives a total orbital flux of 1361 W/m$^2$ and the binary orbital parameters represent a typical CBP host binary within the Kepler CBP systems.}
\label{table:Case_EBM}
\begin{center}
 \begin{tabular}{||c c c c c c c c c c||} 
 \hline
     & $M_A (M_\odot)$ & $M_B (M_\odot)$ & $L_A (L_\odot)$ & $L_B (L_\odot)$ & $T_A $(K) & $T_B$ (K) & $a_{bin}$ (AU) & $e_{bin}$ & $a_{CBP}$ (AU) \\ [0.5ex] 
 \hline\hline
\textit{Case 1}  & 0.84 & 0.84 & 0.5 & 0.5 & 5090 & 5090 & 0.15 & 0.12 & 1 \\ 
 \hline
 \textit{Case 2} & 0.99 & 0.10 & 0.996& 0.004 & 5760 & 1980 & 0.15 & 0.12 & 1\\
 \hline
 \hline
\end{tabular}
\end{center}
\end{table}
We investigate the flux and temperature evolution of two {\it synthetic} systems that are representative of our \textit{Case 1} and \textit{Case 2} categories using an EBM.  The orbital and spectral parameters used for these systems are given in Table \ref{table:Case_EBM}, where we note that the values are chosen so that the CBP receives the same flux (1361 W/m$^2$) and the binary orbital parameters are representative of a typical CBP hosting binary.  Figure \ref{fig:FluxTempDeltaCase1} illustrates the differential latitudinal flux received (left column) and respective surface temperature response (right column) for a \textit{Case 1} CBP with an Earth-like obliquity ($\psi = 23.5^\circ$).  We measure the differential TOA flux $\Delta F$ as the temporal difference in TOA flux between a CBP and an equivalent single star (ESS) system (i.e., $\Delta F = F_{CBP} - F_{ESS}$), where an ESS system is created by setting the planet in the same orbit around a single star with its luminosity equal to the total weighted luminosity of the binary.  The differential surface temperature $\Delta T_s$ is calculated in a similar manner (i.e., $\Delta T_s = T_{CBP}-T_{ESS}$) using the output from the EBM over 5 planetary orbits.  The first orbit is discarded to remove any fluctuations that may occur as the EBM equilibrates. 

In agreement with previous work \citep{May2016}, a change in flux received by the planet creates a change in planet surface temperature, with a time delay because of the nonzero heating/cooling timescale of the planet. Asymmetries about the planet's equator arise due to  the planet's nonzero obliquity and eccentricity, where the CBP's obliquity and eccentricity allow for a higher flux at particular phases of a planetary orbit (i.e., Milankovitch cycles \cite{Milankovitch1941,Deitrick2018}).

In Figure \ref{fig:FluxTempDeltaCase1}a and \ref{fig:FluxTempDeltaCase1}b, the values for \textit{Case 1} from Table \ref{table:Case_EBM} are used and the characteristic fringes illustrate the changing conditions due to the inner binary orbit.  The CBP receives a maximum of 20 W/m$^2$ from the motion of the inner binary, or gyration, (Fig. \ref{fig:FluxTempDeltaCase1}a), which corresponds to an increase in surface temperature by $\sim$2 K (Fig. \ref{fig:FluxTempDeltaCase1}b).  The change in temperature is delayed relative to the change in flux due to the radiative lag of the atmosphere.  After a decrease in the differential flux $\Delta F$, the corresponding change in temperature also decreases.  The decreases are not symmetric due to the asymmetric changes in flux, where an {Earth-analog} CBP is, on average, ~1 K warmer than its ESS counterpart. \cite{May2016} showed a similar effect for Kepler-47b, where the maximum and minimum differential surface temperature is 2 K and -0.5 K, respectively. These asymmetries are not found in \citep{Haqq-Misra2019} because they model the flux from the central star using an analytic periodic function and equate the flux received from the ESS to the mean value of the periodic function. 

The binary semimajor axis (and orbital period) plays a role in the relative distance between the CBP and its hosts stars.  Thus, we explore in Figs. \ref{fig:FluxTempDeltaCase1}c and \ref{fig:FluxTempDeltaCase1}d how a \textit{Case 1} CBP responds (in TOA flux and surface temperature) when the binary orbital period is doubled (or increasing $a_{bin}$ by a factor of $2^{2/3}$).  The longer binary orbital period decreases the frequency of the fringes, where the maximum differential flux increases by a factor of $2^{4/3}$ and is proportional to the square of the change $a_{bin}$.  As a result of the increased flux, the maximum temperature is $\sim$6 K warmer than the ESS configuration.  For longer period binaries, the orbital speed of the binary is slower and the CBP atmosphere has a slightly longer duration at each phase of the binary orbit so that the periods of warming last longer (Fig. \ref{fig:FluxTempDeltaCase1}b vs. Fig \ref{fig:FluxTempDeltaCase1}d).  Both effects work to increase the flux and temperature on the CBP even more than in Fig. \ref{fig:FluxTempDeltaCase1}a and \ref{fig:FluxTempDeltaCase1}b

A larger binary eccentricity shortens the closest distance between the CBP and one of its stars, albeit temporarily.  Thus, we use the conditions in Table \ref{table:Case_EBM} for \textit{Case 1} with double the binary eccentricity $e_{bin}$ and evaluate the EBM in Figs. \ref{fig:FluxTempDeltaCase1}e and \ref{fig:FluxTempDeltaCase1}f.  Although each star comes closer to the CBP, the star also moves along its orbit away from the CBP quickly.  As a result, the increases in differential flux are lower and the surface temperature changes in Fig. \ref{fig:FluxTempDeltaCase1}f are similar to Fig. \ref{fig:FluxTempDeltaCase1}b.  The duration of the highest surface temperature increases are more sporadic.

\subsubsection{Case 2 CBPs} \label{sec:case2_EBM}
For \textit{Case 2}, the luminosity from the secondary star is largely negligible relative to the total TOA flux received by the CBP, but the primary star still orbits at a small distance from the center of mass.  As a result, the gyration effect still occurs and the primary star is not brought as close to the CBP as in \textit{Case 1}.  The variation of the total TOA flux on the CBP is periodic (see Section \ref{sec:methods}), where the surface temperatures can decrease below the ESS case.  Moreover, the maximum differential surface temperature ($\Delta T_s$) is less for \textit{Case 2} than  in \textit{Case 1}.  We demonstrate this effect in Figure \ref{fig:FluxTempDeltaCase2} using the binary and spectral parameters in Table \ref{table:Case_EBM} in a similar experiment as in Section \ref{sec:case1_EBM}.

Figure \ref{fig:FluxTempDeltaCase2}a shows that the differential flux ($\Delta F$) between a \textit{Case 2} CBP and its ESS configuration is more sinusoidal.  The differential surface temperature (Fig. \ref{fig:FluxTempDeltaCase2}b) exhibits features that correlate with the differential flux, but are delayed due to the radiative lag of the atmosphere.  Overall, the maximum surface temperature is only $\sim${0.5} K warmer than its ESS counterpart, where most latitudes experience a $\sim${0.3} K difference in surface temperature.

In Section \ref{sec:case1_EBM} we doubled the binary orbital period to induce a larger differential flux, but the increase in this case is {less than} half as much compared to \textit{Case 1} (Fig. \ref{fig:FluxTempDeltaCase2}c vs. \ref{fig:FluxTempDeltaCase1}c).  The resulting differential surface temperature (Fig. \ref{fig:FluxTempDeltaCase2}d) is also not as large, but roughly double in magnitude relative to Fig. \ref{fig:FluxTempDeltaCase2}b.  If we double the binary eccentricity, the differential flux (Fig. \ref{fig:FluxTempDeltaCase2}e) is similar to Fig. \ref{fig:FluxTempDeltaCase2}a.  The periods when the differential surface temperatures are lower than the ESS configuration (Fig. \ref{fig:FluxTempDeltaCase2}f) are longer in duration for a higher binary eccentricity, which makes \textit{Case 2} CBPs more similar to the ESS configuration.

As shown in Fig \ref{fig:FluxVarMu}d, increasing binary eccentricity increases the maximum flux and decreases the minimum flux received by the CBP, creating a larger range in fluxes received. In \textit{Case 1}, the stars have comparable luminosities and any increase in flux received from one star due to the binary orbit is countered by a corresponding decrease in flux from the other star. Thus, when binary eccentricity is increased in \textit{Case 1}, the total flux received by the CBP at any point does not change significantly enough to drive a temperature change. In \textit{Case 2}, the secondary is much dimmer than the primary and doesn't counter the changes in flux from the primary. Thus, eccentricity changes to the binary in \textit{Case 2} cause more significant variability in the temperatures for an {Earth-analog} CBP. 

In every scenario presented in Figs. \ref{fig:FluxTempDeltaCase1} and \ref{fig:FluxTempDeltaCase2}, we find that the surface temperatures on an {Earth-analog} CBP are typically warmer than the equivalent single-star configuration.  The maximum difference is typically not that large, similar to previous works \citep{May2016, Haqq-Misra2019}, and more specific CBP configurations optimized for a larger differential flux could induce larger temperatures differences than in these cases.

\subsubsection{Flux and Temperatures on Known HZ CBPs}
We apply our analysis for the differential flux ($\Delta F$) and surface temperature ($\Delta T_s$) using parameters from four of the known Kepler CBPs (Kepler-16b, Kepler-47c, Kepler-453b, and Kepler-1647b) that lie within their respective conservative habitable zones \citep{Haghighipour2013}, but we replace the gas giants with {Earth analogs}.  The orbital parameters of these systems are given in Table \ref{table:Plot8CBPValues} along with their flux variation $F_V$.  The $F_V$ values are similar for each of the Kepler CBPs because the planetary eccentricity $e_{CBP}$ induces a variation over a planetary orbit in addition to the gyration of the inner binary.  If the orbits of the more eccentric Kepler CBPs (47c, 453b, and 1647b) were circular, then the $F_V$ values would be 0.09, 0.16, and 0.06, respectively.  To remove the variability due to planetary eccentricity, we present results assuming circular, coplanar planetary orbits and the actual flux variation of these systems will be enhanced using the planetary parameters.

For simplicity, we calculate the differential flux and surface temperature evolution for an {Earth-analog} with $0^\circ$ of obliquity in Figure \ref{fig:Kep_CBP_time}.  Similar to our investigation of the general cases (Sections \ref{sec:case1_EBM} and \ref{sec:case2_EBM}), we omit the first full planetary orbit as the atmosphere equilibrates.  A Kepler-16b analog can receive up to {$\sim$35 W/m$^2$} additional flux relative to its ESS configuration, which occurs when the planet and primary star are at their shortest distance.  The differential surface temperature for an {Earth-analog} Kepler 16b increases by {1.2} K in response to the additional flux and is delayed by $\sim$0.15 orbits due to the radiative lag of the atmosphere.  Throughout the planetary orbit, the binary shows a range of relative phases that allows the differential temperature to relax back towards a state more similar to its ESS counterpart.  However the differential temperature remains positive (CBP is warmer than the ESS planet) for an overwhelming majority of an orbit.

An {Earth-analog} to Kepler-47c is more symmetric for its differential flux as compared to the ESS configuration, but it is always warmer than the respective ESS case.  This is, in part, due to the relatively short binary orbital period where changes in flux are on the same order as the atmospheric equilibration timescale.  Kepler-453AB, on the other hand, has a longer binary orbital period, which allows the atmosphere of an {Earth-analog} orbiting it more time to reach equilibrium once the flux changes.  As a result, the maximum differential flux reaches $\sim$50 W/m$^2$ with a nearly 3K increase in the corresponding differential surface temperature.  Kepler-1647AB has a short period ($\sim$11 days) inner binary, where the planetary orbital period is extremely long ($\sim$1100 days).  As a consequence, the differential flux is largely symmetric (similar to Kepler-47c) and the differential surface temperature on an {Earth-analog} is more similar to its ESS configuration than of the other Kepler CBPs.  An extreme CBP that lies near its stability limit and inner edge of its HZ experiences at least 4 epochs of extreme differential flux ($\sim$180 W/m$^2$), which correlates with a 10 - 15 K increase in the differential surface temperature.  The range in surface temperature arises from the movement of the CBP along its orbit and the binary presents a different phase relative to the CBP.

Many previous studies \citep{Armstrong2014,Kane2017,Deitrick2018,Quarles2020b} have shown significant latitudinal changes in flux with respect to the planet's obliquity $\psi$.  Figure \ref{fig:Kep_CBP_max} shows the maximum differential flux and surface temperature (color-coded) using {Earth-analogs} in place of the four known habitable zone CBPs and one hypothetical system with obliquities ranging from 0$^\circ$ to $180^\circ$.  We also provide the base-10 logarithm of the ratio $\gamma$, where $\gamma$ represents the ratio of the maximum differential flux and surface temperature during an orbit (i.e., max $\Delta T_s$/ max $\Delta F$).  Physically, the ratio $\gamma$ indicates the magnitude of the increase in surface temperature relative to the additional flux introduced by the motion of the inner binary (i.e., gyration).  For our Kepler-16b analogs (top row, Fig. \ref{fig:Kep_CBP_max}), the greatest differential flux occurs at 90$^\circ$ of obliquity resulting in a {6} K increase in the differential surface temperature.  Through $\log_{10}\;\gamma$, we find that the thermal atmospheric response varies with obliquity and latitude, which is also large proportional to the input of additional flux from the changing binary orbit relative to the ESS configuration.

The {Earth-analogs} in Kepler-47c and Kepler-453b exhibit similar trends as Kepler-16b for the maximum differential flux, where the magnitudes vary due to the spectral properties of the host stars.  But, the surface temperature on these planets begins above the freezing temperature for sea water assumed in our EBM, in contrast to Kepler-16b (see Figure \ref{fig:Kep_CBP_med}).  Therefore, the planets will encounter a critical obliquity range (28$^\circ$ -- 152$^\circ$, Kepler-47c; 57$^\circ$ -- 125$^\circ$, Kepler-453b)  for which the poles start to receive more flux than the equator and an ice belt can form \citep{Kilic2018}.  An ice belt changes the average albedo of the planet and can lead to more drastic cooling.  The differential surface temperature panels (middle column, Figure \ref{fig:Kep_CBP_max}) for Kepler-47c and Kepler-453b show the polar regions are more similar to their respective ESS configurations between the critical obliquity values (vertical ridges in red).  This effect appears more clearly through $\log_{10}\;\gamma$, where the atmospheric response is significantly lower between the critical obliquities.  The maximum differential flux and surface temperature for Kepler-1647b is much more similar to Kepler-16b, but the overall degree is much less due to the orbital architecture.  The maximum differential surface temperature can increase by $\sim$1 K in Kepler-1647b, but only at high obliquity and in the polar regions.  Kepler-1647b begins with a lower global surface temperature so that there is not a large shift in albedo with an increased obliquity. 

A more extreme CBP configuration is possible (although not yet discovered) that has a semimajor axis $a_{CBP}$ near its stability limit and is also at the inner edge of its HZ.  We explore this hypothetical case (Extreme CBP) in Fig. \ref{fig:Kep_CBP_max} and find that such a planet could experience up to a {$\sim$600 W/m$^2$} increase in its differential flux.  As a result, the differential surface temperature increases by {$\sim$23 K}.  In this case, the gyration of the binary largely prevents ice belts from forming at high obliquity and thus the global albedo remains quite low.  The atmospheric response in $\log_{10}\;\gamma$ is typically {between -1 and -2 (or $\gamma $ between 0.01 and 0.1 )}, which means every 10 W/m$^2$ increase in differential flux correlates with a {0.1-1} K increase in the differential surface temperature.

The median surface temperature ($T_{med}$) for each CBP changes as a function of the assumed obliquity (Figure \ref{fig:Kep_CBP_med}).  {Earth-analogs} in place of Kepler-16b and Kepler-1647b follow their respective ESS configurations due to their surface temperatures beginning below our freezing point (263 K) so that the global albedo remains the same across obliquities.  Kepler-47c and Kepler-453b begin at a warmer median surface temperatures $T_{med}$ until the equator receives less flux due to the obliquity and the gyration from the binary.  An ice belt can potentially form, which drastically changes the global albedo through an ice-albedo feedback.  This process can be interrupted if the flux variations from the binary gyration are more extreme as is the case for hypothetical extreme CBP (Fig. \ref{fig:Kep_CBP_med}; bottom row).  The extent of cooling due to nearly 90$^\circ$ obliquity for the Kepler CBPs is also mitigated due to the flux variation from the inner binary.  Although, it is typically to a lower degree than in our extreme CBP scenario.

\subsection{Binary Eclipses}
Our analysis has ignored the binary eclipses that affect coplanar CBPs, where the total flux received by the CBP is temporarily and drastically reduced. Figure \ref{fig:FluxLatEclipses} shows the daily averaged flux for Kepler-16b over one planetary orbit (assuming Earth-like obliquity), where the binary eclipses are seen as short vertical strips along the fringe pattern of the binary.  The duration of binary eclipses depends on the orbital velocity of the host stars, where eclipses of Kepler-16AB have the longest duration because of its large separation $a_{bin}$ and consequently the longest orbital period of the Kepler CBPs.  Additionally, the eclipse duration  depends specifically on which star is being occulted. Nonetheless, we find using numerical integrations in \texttt{Rebound} that the eclipses typically last between $\soom$ 7 -- 11 hours, which is much less than the equilibration timescale of the atmosphere and will not be significant for the long-term change in TOA flux or surface temperature on CBPs.  Although eclipsing binaries as seen from the CBP are likely rare, CBPs in general may temporarily experience binary eclipses due to the secular nodal precession, which is most drastically seen for Kepler-413b \citep{Kostov2016Kep1647Info} and Kepler-1661b \citep{Socia2020Kep1661Info}.    

\section{Summary \& Discussions} \label{sec:summary}
When examining the habitability of a circumbinary planet (CBP), one must not only consider the average stellar flux received by the CBP, but also account for the orbital motions and intrinsic characteristics of the inner binary \citep{Eggl2014}. Because each star of the binary has its own motion around the center of mass of the system (i.e., gyration), the flux received by the CBP can vary substantially during a single planetary orbit. In this work, we present an analysis of the top-of-atmosphere (TOA) flux and surface temperature experienced by an {Earth-analog} CBP in orbit around two main-sequence stars, with an emphasis on parameters from the currently known CBPs.

With respect to the TOA flux received by a CBP, we find two types of configurations, as shown in Fig. \ref{fig:FluxMaxMinCBP}. If the secondary star's luminosity is comparable to that of the primary star (\textit{Case 1}: $L_B \soom L_A$), it makes a non-negligible contribution to the CBP's surface flux. In \textit{Case 1}, the flux received by the CBP varies on two timescales -- one-half of the binary orbital period and the CBP orbital period. If the weighted luminosity of the secondary star is much lower than the weighted luminosity of the primary star (\textit{Case 2}: $L_B << L_A$), its contribution to flux doesn't affect the TOA flux and CBP flux variation is mainly driven by the periodically varying distance between the CBP and the primary stellar component. In contrast to \textit{Case 1}, the flux received by the CBP in \textit{Case 2} varies on the binary orbital period timescale. We present an analytical expression (equation \ref{eq:BorderEquation}) for the luminosity ratio of the border between \textit{Case 1} and \textit{Case 2}, in terms of the dynamical mass ratio $\mu$, the semimajor axis ratio $a_{CBP}/a_{bin}$, and the binary eccentricity $e_{bin}$. The phase region corresponding to \textit{Case 2} is much larger than than of \textit{Case 1} (see Fig. \ref{fig:FluxVarMu}d), and we expect \textit{Case 2} configurations to more commonly appear in observed systems. This is supported by the known Kepler CBPs, where only two (Kepler-34b and Kepler-35b) are \textit{Case 1} configurations out of the dozen known CBPs. Moreover, the fraction of binaries with a Solar-Type primary and a low-mass secondary is expected to be high \citep{Raghavan2010,Moe2017}.

We characterize the TOA flux variations experienced by a CBP in terms of the dynamical parameters of the system. For a given primary mass, the flux variation peaks at $\mu$ $\sim$0.3 and varying the primary mass can shift this peak slightly, where the flux variation peaks at $\mu = 0.24$, $\mu = 0.29$, and $\mu = 0.33$ for an M-dwarf, K-dwarf, and G-dwarf primary, respectively (Fig. \ref{fig:FluxVarMu}a and \ref{fig:FluxVarMu}b). As the semimajor axis ratio $a_{CBP}/a_{bin}$ increases, the binary more closely resembles a single star and the flux variation decreases (Fig. \ref{fig:FluxVarMu}c). When the binary eccentricity increases, the maximum flux on the CBP increases and the minimum flux decreases, uniformly increasing the flux variation (Fig. \ref{fig:FluxVarMu}d). We find that these trends are independent of whether the CBP is near the inner HZ boundary (runaway limit) or the outer HZ boundary (greenhouse limit).  Using a 1D Longitudinal energy balance model (EBM), we evaluate the differential flux ($\Delta F = F_{CBP}-F_{ESS}$) and surface temperatures ($\Delta T_s = T_{CBP}-T_{ESS}$) for \textit{Case 1} and \textit{Case 2} {Earth-analog} CBPs relative to their equivalent single-star (ESS) configurations. We find that {Earth-analog} CBPs are slightly warmer than their ESS configuration. As shown in Fig. \ref{fig:FluxTempDeltaCase1} \& \ref{fig:FluxTempDeltaCase2}, Earth-analog \textit{Case 1} CBPs can be 6 K warmer than the ESS configuration, and Earth-analog \textit{Case 2} CBPs can be 2 K warmer. Because the stellar mass and luminosity are less evenly distributed in \textit{Case 2}, changes in the binary orbital parameters drive larger temperature changes compared to \textit{Case 1} CBPs. \cite{May2016} concluded that the surface temperature for CBPs are very similar to their respective ESS configurations, where we find agreement for low obliquities.  A key difference between our EBM and their model is that our albedo changes as the temperature drops below 263 K, which produces a feedback.  \cite{Haqq-Misra2019} found that CBPs can experience significant differences in surface temperature from the ESS configuration when the CBP's surface is assumed to have more land, where our results agree in the case of an aquaworld. In contrast, \cite{Haqq-Misra2019} found a more symmetric distribution of differential surface temperatures. This discrepancy is due to the way they model the flux received from the central binary.  Overall, each model agrees that CBPs can experience warmer surface temperatures compared with the ESS configuration.

In addition, we evaluate the expected differential flux and surface temperatures for {Earth-analogs} using parameters from the four known habitable zone CBPs and one extreme hypothetical case over one planetary orbit (Fig. \ref{fig:Kep_CBP_time}). Because a longer binary period allows the CBP atmosphere more time to equilibriate to changes in TOA flux, CBPs with relatively short period host binaries (Kepler-47c and Kepler-1647b) have a relatively low $\Delta T$ than CBPs with longer binary periods (Kepler-16b, Kepler-453b, and Extreme CBP; Fig. \ref{fig:Kep_CBP_time}). We evaluate the maximum differential flux, maximum differential temperature, and their ratio for a wide range of obliquity in Fig. \ref{fig:Kep_CBP_max}. At higher obliquities, as the polar latitudes of CBPs receive more direct flux than the equator, ice belts begin to form near the equator, which can drastically change the global CBP albedo. We find that the change in albedo can drive a drastic shift in differential temperature on {Earth-analog} CBPS in systems like Kepler-47c and Kepler-453b.  CBPs with high obliquity ($\psi\sim 90^\circ$) typically have a larger differential temperature at the poles due to direct radiation, where diffusion acts to increase the differential temperature at lower latitudes.  The ice-albedo feedback can be counterbalanced by the gyration of the binary orbit when the CBP begins near its stability limit and at the inner edge of its HZ due to the more drastic periodic increases in flux (Figs. \ref{fig:Kep_CBP_max} and \ref{fig:Kep_CBP_med}).

Our analysis applies to {Earth-analog} CBPs that are nearly coplanar with their host binary. While it is expected that most CBPs have a low mutual inclination with their host stars, it is possible that more inclined CBPs exist \citep{Li2016}. A CBP's orbit can be stable at a smaller semimajor axis at higher inclinations \citep{Doolin2011}. As a consequence, highly inclined CBPs could see larger flux variations. This effect would be most prominent for retrograde CBPs, which can be significantly closer to the central binary than prograde CBPs and might be overall less habitable than prograde CBPs. In Kepler-47 \citep{Orosz2012Kep47bc} and potentially Kepler-1647 \citep{Hong2019}, it is possible that multiple CBPs exist in orbit around the same inner binary. The work in \cite{Sutherland2019} shows that the instabilities in mean motion resonances, which are expected in such multiplanet circumbinary environments, could create a more nuanced analysis of CBP habitability. In addition, the work presented in this paper can be extended by using more complex and long-term climate models to derive detailed information about the habitability and temperature on the surface of CBPs. As in \cite{Haqq-Misra2019}, accounting for surface topography and land distribution could create an extended investigation of CBP temperature and habitability.

{\section*{Acknowledgements}
The authors thank the anonymous reviewer whose comments greatly improved the content and clarity of the manuscript. N.H. acknowledges support from NASA XRP through grant number 80NSSC18K0519. G. L. acknowledges support from NASA ATP through grant number 80NSSC20K0522. }

\section*{Data availability}
The data underlying this article will be shared on reasonable request to the corresponding author.

\bibliographystyle{mnras}
\bibliography{biblio}

\begin{thebibliography}{}
\makeatletter
\relax
\def\mn@urlcharsother{\let\do\@makeother \do\$\do\&\do\#\do\^\do\_\do\%\do\~}
\def\mn@doi{\begingroup\mn@urlcharsother \@ifnextchar [ {\mn@doi@}
  {\mn@doi@[]}}
\def\mn@doi@[#1]#2{\def\@tempa{#1}\ifx\@tempa\@empty \href
  {http://dx.doi.org/#2} {doi:#2}\else \href {http://dx.doi.org/#2} {#1}\fi
  \endgroup}
\def\mn@eprint#1#2{\mn@eprint@#1:#2::\@nil}
\def\mn@eprint@arXiv#1{\href {http://arxiv.org/abs/#1} {{\tt arXiv:#1}}}
\def\mn@eprint@dblp#1{\href {http://dblp.uni-trier.de/rec/bibtex/#1.xml}
  {dblp:#1}}
\def\mn@eprint@#1:#2:#3:#4\@nil{\def\@tempa {#1}\def\@tempb {#2}\def\@tempc
  {#3}\ifx \@tempc \@empty \let \@tempc \@tempb \let \@tempb \@tempa \fi \ifx
  \@tempb \@empty \def\@tempb {arXiv}\fi \@ifundefined
  {mn@eprint@\@tempb}{\@tempb:\@tempc}{\expandafter \expandafter \csname
  mn@eprint@\@tempb\endcsname \expandafter{\@tempc}}}

\bibitem[\protect\citeauthoryear{{Armstrong}, {Barnes}, {Domagal-Goldman},
  {Breiner}, {Quinn}  \& {Meadows}}{{Armstrong} et~al.}{2014}]{Armstrong2014}
{Armstrong} J.~C.,  {Barnes} R.,  {Domagal-Goldman} S.,  {Breiner} J.,  {Quinn}
  T.~R.,   {Meadows} V.~S.,  2014, \mn@doi [Astrobiology]
  {10.1089/ast.2013.1129}, \href
  {https://ui.adsabs.harvard.edu/abs/2014AsBio..14..277A} {14, 277}

\bibitem[\protect\citeauthoryear{{Borucki} et~al.,}{{Borucki}
  et~al.}{2010}]{Borucki2010}
{Borucki} W.~J.,  et~al., 2010, \mn@doi [Science] {10.1126/science.1185402},
  \href {https://ui.adsabs.harvard.edu/abs/2010Sci...327..977B} {327, 977}

\bibitem[\protect\citeauthoryear{{Cukier}, {Kopparapu}, {Kane}, {Welsh},
  {Wolf}, {Kostov}  \& {Haqq-Misra}}{{Cukier} et~al.}{2019}]{Cukier2019}
{Cukier} W.,  {Kopparapu} R.~k.,  {Kane} S.~R.,  {Welsh} W.,  {Wolf} E.,
  {Kostov} V.,   {Haqq-Misra} J.,  2019, \mn@doi [\pasp]
  {10.1088/1538-3873/ab50cb}, \href
  {https://ui.adsabs.harvard.edu/abs/2019PASP..131l4402C} {131, 124402}

\bibitem[\protect\citeauthoryear{{Deitrick}, {Barnes}, {Quinn}, {Armstrong},
  {Charnay}  \& {Wilhelm}}{{Deitrick} et~al.}{2018}]{Deitrick2018}
{Deitrick} R.,  {Barnes} R.,  {Quinn} T.~R.,  {Armstrong} J.,  {Charnay} B.,
  {Wilhelm} C.,  2018, \mn@doi [\aj] {10.3847/1538-3881/aaa301}, \href
  {https://ui.adsabs.harvard.edu/abs/2018AJ....155...60D} {155, 60}

\bibitem[\protect\citeauthoryear{Doolin \& Blundell}{Doolin \&
  Blundell}{2011}]{Doolin2011}
Doolin S.,  Blundell K.~M.,  2011, \mn@doi [Monthly Notices of the Royal
  Astronomical Society] {10.1111/j.1365-2966.2011.19657.x}, 418, 2656

\bibitem[\protect\citeauthoryear{Doyle et~al.,}{Doyle
  et~al.}{2011}]{Doyle2011Kep16Info}
Doyle L.~R.,  et~al., 2011, \mn@doi [Science] {10.1126/science.1210923}, 333,
  1602

\bibitem[\protect\citeauthoryear{{Dvorak}}{{Dvorak}}{1982}]{Dvorak1982}
{Dvorak} R.,  1982, Oesterreichische Akademie Wissenschaften Mathematisch
  naturwissenschaftliche Klasse Sitzungsberichte Abteilung, \href
  {https://ui.adsabs.harvard.edu/abs/1982OAWMN.191..423D} {191, 423}

\bibitem[\protect\citeauthoryear{{Dvorak}}{{Dvorak}}{1986}]{Dvorak1986}
{Dvorak} R.,  1986, \aap, \href
  {https://ui.adsabs.harvard.edu/abs/1986A&A...167..379D} {167, 379}

\bibitem[\protect\citeauthoryear{{Eggl}, {Georgakarakos}  \&
  {Pilat-Lohinger}}{{Eggl} et~al.}{2014}]{Eggl2014}
{Eggl} S.,  {Georgakarakos} N.,   {Pilat-Lohinger} E.,  2014, in Complex
  Planetary Systems, Proceedings of the International Astronomical Union. pp
  53--57 (\mn@eprint {arXiv} {1412.1118}), \mn@doi{10.1017/S1743921314007820}

\bibitem[\protect\citeauthoryear{{Forgan}}{{Forgan}}{2014}]{Forgan2014}
{Forgan} D.,  2014, \mn@doi [\mnras] {10.1093/mnras/stt1964}, \href
  {https://ui.adsabs.harvard.edu/abs/2014MNRAS.437.1352F} {437, 1352}

\bibitem[\protect\citeauthoryear{{Forgan}, {Mead}, {Cockell}  \&
  {Raven}}{{Forgan} et~al.}{2015}]{Forgan2015}
{Forgan} D.~H.,  {Mead} A.,  {Cockell} C.~S.,   {Raven} J.~A.,  2015, \mn@doi
  [International Journal of Astrobiology] {10.1017/S147355041400041X}, \href
  {https://ui.adsabs.harvard.edu/abs/2015IJAsB..14..465F} {14, 465}

\bibitem[\protect\citeauthoryear{{Haghighipour} \&
  {Kaltenegger}}{{Haghighipour} \& {Kaltenegger}}{2013}]{Haghighipour2013}
{Haghighipour} N.,  {Kaltenegger} L.,  2013, \mn@doi [\apj]
  {10.1088/0004-637X/777/2/166}, \href
  {http://adsabs.harvard.edu/abs/2013ApJ...777..166H} {777, 166}

\bibitem[\protect\citeauthoryear{{Haqq-Misra}, {Wolf}, {Welsh}, {Kopparapu},
  {Kostov}  \& {Kane}}{{Haqq-Misra} et~al.}{2019}]{Haqq-Misra2019}
{Haqq-Misra} J.,  {Wolf} E.~T.,  {Welsh} W.~F.,  {Kopparapu} R.~K.,  {Kostov}
  V.,   {Kane} S.~R.,  2019, \mn@doi [Journal of Geophysical Research
  (Planets)] {10.1029/2019JE006222}, \href
  {https://ui.adsabs.harvard.edu/abs/2019JGRE..124.3231H} {124, 3231}

\bibitem[\protect\citeauthoryear{{Holman} \& {Wiegert}}{{Holman} \&
  {Wiegert}}{1999}]{Holman1999}
{Holman} M.~J.,  {Wiegert} P.~A.,  1999, \mn@doi [\aj] {10.1086/300695}, \href
  {https://ui.adsabs.harvard.edu/abs/1999AJ....117..621H} {117, 621}

\bibitem[\protect\citeauthoryear{{Hong}, {Quarles}, {Li}  \& {Orosz}}{{Hong}
  et~al.}{2019}]{Hong2019}
{Hong} Z.,  {Quarles} B.,  {Li} G.,   {Orosz} J.~A.,  2019, \mn@doi [\aj]
  {10.3847/1538-3881/ab2127}, \href
  {https://ui.adsabs.harvard.edu/abs/2019AJ....158....8H} {158, 8}

\bibitem[\protect\citeauthoryear{{Hurowitz} et~al.,}{{Hurowitz}
  et~al.}{2017}]{Hurowitz2017}
{Hurowitz} J.~A.,  et~al., 2017, \mn@doi [Science] {10.1126/science.aah6849},
  \href {https://ui.adsabs.harvard.edu/abs/2017Sci...356.6849H} {356, aah6849}

\bibitem[\protect\citeauthoryear{{Kane} \& {Hinkel}}{{Kane} \&
  {Hinkel}}{2013}]{Kane2013}
{Kane} S.~R.,  {Hinkel} N.~R.,  2013, \mn@doi [\apj]
  {10.1088/0004-637X/762/1/7}, \href
  {https://ui.adsabs.harvard.edu/abs/2013ApJ...762....7K} {762, 7}

\bibitem[\protect\citeauthoryear{{Kane} \& {Torres}}{{Kane} \&
  {Torres}}{2017}]{Kane2017}
{Kane} S.~R.,  {Torres} S.~M.,  2017, \mn@doi [\aj] {10.3847/1538-3881/aa8fce},
  \href {https://ui.adsabs.harvard.edu/abs/2017AJ....154..204K} {154, 204}

\bibitem[\protect\citeauthoryear{{Kasting}}{{Kasting}}{1991}]{KASTING1991}
{Kasting} J.~F.,  1991, \mn@doi [\icarus] {10.1016/0019-1035(91)90137-I}, \href
  {https://ui.adsabs.harvard.edu/abs/1991Icar...94....1K} {94, 1}

\bibitem[\protect\citeauthoryear{{Kasting} \& {Pollack}}{{Kasting} \&
  {Pollack}}{1983}]{Kasting1983}
{Kasting} J.~F.,  {Pollack} J.~B.,  1983, \mn@doi [\icarus]
  {10.1016/0019-1035(83)90212-9}, \href
  {https://ui.adsabs.harvard.edu/abs/1983Icar...53..479K} {53, 479}

\bibitem[\protect\citeauthoryear{{Kasting}, {Whitmire}  \&
  {Reynolds}}{{Kasting} et~al.}{1993}]{KASTING1993}
{Kasting} J.~F.,  {Whitmire} D.~P.,   {Reynolds} R.~T.,  1993, \mn@doi
  [\icarus] {10.1006/icar.1993.1010}, \href
  {https://ui.adsabs.harvard.edu/abs/1993Icar..101..108K} {101, 108}

\bibitem[\protect\citeauthoryear{{Kilic}, {Lunkeit}, {Raible}  \&
  {Stocker}}{{Kilic} et~al.}{2018}]{Kilic2018}
{Kilic} C.,  {Lunkeit} F.,  {Raible} C.~C.,   {Stocker} T.~F.,  2018, \mn@doi
  [\apj] {10.3847/1538-4357/aad5eb}, \href
  {https://ui.adsabs.harvard.edu/abs/2018ApJ...864..106K} {864, 106}

\bibitem[\protect\citeauthoryear{{Kopparapu} et~al.,}{{Kopparapu}
  et~al.}{2013}]{Kopparapu2013}
{Kopparapu} R.~K.,  et~al., 2013, \mn@doi [\apj] {10.1088/0004-637X/765/2/131},
  \href {https://ui.adsabs.harvard.edu/abs/2013ApJ...765..131K} {765, 131}

\bibitem[\protect\citeauthoryear{{Kopparapu}, {Ramirez}, {SchottelKotte},
  {Kasting}, {Domagal-Goldman}  \& {Eymet}}{{Kopparapu}
  et~al.}{2014}]{Kopparapu2014}
{Kopparapu} R.~K.,  {Ramirez} R.~M.,  {SchottelKotte} J.,  {Kasting} J.~F.,
  {Domagal-Goldman} S.,   {Eymet} V.,  2014, \mn@doi [\apjl]
  {10.1088/2041-8205/787/2/L29}, \href
  {https://ui.adsabs.harvard.edu/abs/2014ApJ...787L..29K} {787, L29}

\bibitem[\protect\citeauthoryear{{Kostov} et~al.,}{{Kostov}
  et~al.}{2014}]{Kostov2014Kep413Info}
{Kostov} V.~B.,  et~al., 2014, \mn@doi [\apj] {10.1088/0004-637X/784/1/14},
  \href {https://ui.adsabs.harvard.edu/abs/2014ApJ...784...14K} {784, 14}

\bibitem[\protect\citeauthoryear{{Kostov} et~al.,}{{Kostov}
  et~al.}{2016}]{Kostov2016Kep1647Info}
{Kostov} V.~B.,  et~al., 2016, \mn@doi [\apj] {10.3847/0004-637X/827/1/86},
  \href {https://ui.adsabs.harvard.edu/abs/2016ApJ...827...86K} {827, 86}

\bibitem[\protect\citeauthoryear{{Kostov} et~al.,}{{Kostov}
  et~al.}{2020}]{Kostov2020TOI1338Info}
{Kostov} V.~B.,  et~al., 2020, \mn@doi [\aj] {10.3847/1538-3881/ab8a48}, \href
  {https://ui.adsabs.harvard.edu/abs/2020AJ....159..253K} {159, 253}

\bibitem[\protect\citeauthoryear{{Li}, {Holman}  \& {Tao}}{{Li}
  et~al.}{2016}]{Li2016}
{Li} G.,  {Holman} M.~J.,   {Tao} M.,  2016, \mn@doi [\apj]
  {10.3847/0004-637X/831/1/96}, \href
  {https://ui.adsabs.harvard.edu/abs/2016ApJ...831...96L} {831, 96}

\bibitem[\protect\citeauthoryear{Mason, Zuluaga, Clark  \&
  Cuartas-Restrepo}{Mason et~al.}{2013}]{Mason2013}
Mason P.~A.,  Zuluaga J.~I.,  Clark J.~M.,   Cuartas-Restrepo P.~A.,  2013,
  \mn@doi [The Astrophysical Journal] {10.1088/2041-8205/774/2/l26}, 774, L26

\bibitem[\protect\citeauthoryear{{Mason}, {Zuluaga}, {Cuartas-Restrepo}  \&
  {Clark}}{{Mason} et~al.}{2015}]{Mason2015}
{Mason} P.~A.,  {Zuluaga} J.~I.,  {Cuartas-Restrepo} P.~A.,   {Clark} J.~M.,
  2015, \mn@doi [International Journal of Astrobiology]
  {10.1017/S1473550414000342}, \href
  {https://ui.adsabs.harvard.edu/abs/2015IJAsB..14..391M} {14, 391}

\bibitem[\protect\citeauthoryear{May \& Rauscher}{May \&
  Rauscher}{2016}]{May2016}
May E.~M.,  Rauscher E.,  2016, \mn@doi [The Astrophysical Journal]
  {10.3847/0004-637x/826/2/225}, 826, 225

\bibitem[\protect\citeauthoryear{Milankovitch}{Milankovitch}{1941}]{Milankovitch1941}
Milankovitch M.~K.,  1941, Royal Serbian Academy Special Publication, 133, 1

\bibitem[\protect\citeauthoryear{{Moe} \& {Di Stefano}}{{Moe} \& {Di
  Stefano}}{2017}]{Moe2017}
{Moe} M.,  {Di Stefano} R.,  2017, \mn@doi [\apjs] {10.3847/1538-4365/aa6fb6},
  \href {https://ui.adsabs.harvard.edu/abs/2017ApJS..230...15M} {230, 15}

\bibitem[\protect\citeauthoryear{{Murray} \& {Dermott}}{{Murray} \&
  {Dermott}}{1999}]{Murray1999}
{Murray} C.~D.,  {Dermott} S.~F.,  1999, {Solar system dynamics}

\bibitem[\protect\citeauthoryear{{Orosz} et~al.,}{{Orosz}
  et~al.}{2012a}]{Orosz2012Kep47bc}
{Orosz} J.~A.,  et~al., 2012a, \mn@doi [Science] {10.1126/science.1228380},
  \href {https://ui.adsabs.harvard.edu/abs/2012Sci...337.1511O} {337, 1511}

\bibitem[\protect\citeauthoryear{{Orosz} et~al.,}{{Orosz}
  et~al.}{2012b}]{Orosz2012Kep38Info}
{Orosz} J.~A.,  et~al., 2012b, \mn@doi [\apj] {10.1088/0004-637X/758/2/87},
  \href {https://ui.adsabs.harvard.edu/abs/2012ApJ...758...87O} {758, 87}

\bibitem[\protect\citeauthoryear{{Orosz} et~al.,}{{Orosz}
  et~al.}{2019}]{Orosz2019Kep47dInfo}
{Orosz} J.~A.,  et~al., 2019, \mn@doi [\aj] {10.3847/1538-3881/ab0ca0}, \href
  {https://ui.adsabs.harvard.edu/abs/2019AJ....157..174O} {157, 174}

\bibitem[\protect\citeauthoryear{Popp \& Eggl}{Popp \& Eggl}{2017}]{Popp2017}
Popp M.,  Eggl S.,  2017, Nature Communications, 8, 14957

\bibitem[\protect\citeauthoryear{{Quarles}, {Musielak}  \& {Cuntz}}{{Quarles}
  et~al.}{2012}]{Quarles2012}
{Quarles} B.,  {Musielak} Z.~E.,   {Cuntz} M.,  2012, \mn@doi [\apj]
  {10.1088/0004-637X/750/1/14}, \href
  {https://ui.adsabs.harvard.edu/abs/2012ApJ...750...14Q} {750, 14}

\bibitem[\protect\citeauthoryear{{Quarles}, {Satyal}, {Kostov}, {Kaib}  \&
  {Haghighipour}}{{Quarles} et~al.}{2018}]{Quarles2018}
{Quarles} B.,  {Satyal} S.,  {Kostov} V.,  {Kaib} N.,   {Haghighipour} N.,
  2018, \mn@doi [\apj] {10.3847/1538-4357/aab264}, \href
  {http://adsabs.harvard.edu/abs/2018ApJ...856..150Q} {856, 150}

\bibitem[\protect\citeauthoryear{{Quarles}, {Barnes}, {Lissauer}  \&
  {Chambers}}{{Quarles} et~al.}{2020a}]{Quarles2020b}
{Quarles} B.,  {Barnes} J.~W.,  {Lissauer} J.~J.,   {Chambers} J.,  2020a,
  \mn@doi [Astrobiology] {10.1089/ast.2018.1932}, \href
  {https://ui.adsabs.harvard.edu/abs/2020AsBio..20...73Q} {20, 73}

\bibitem[\protect\citeauthoryear{{Quarles}, {Li}, {Kostov}  \&
  {Haghighipour}}{{Quarles} et~al.}{2020b}]{Quarles2020}
{Quarles} B.,  {Li} G.,  {Kostov} V.,   {Haghighipour} N.,  2020b, \mn@doi
  [\aj] {10.3847/1538-3881/ab64fa}, \href
  {https://ui.adsabs.harvard.edu/abs/2020AJ....159...80Q} {159, 80}

\bibitem[\protect\citeauthoryear{{Rabl} \& {Dvorak}}{{Rabl} \&
  {Dvorak}}{1988}]{Rabl1988}
{Rabl} G.,  {Dvorak} R.,  1988, \aap, \href
  {https://ui.adsabs.harvard.edu/abs/1988A&A...191..385R} {191, 385}

\bibitem[\protect\citeauthoryear{{Raghavan} et~al.,}{{Raghavan}
  et~al.}{2010}]{Raghavan2010}
{Raghavan} D.,  et~al., 2010, \mn@doi [\apjs] {10.1088/0067-0049/190/1/1},
  \href {https://ui.adsabs.harvard.edu/abs/2010ApJS..190....1R} {190, 1}

\bibitem[\protect\citeauthoryear{{Rein} \& {Liu}}{{Rein} \&
  {Liu}}{2012}]{Rein2012}
{Rein} H.,  {Liu} S.-F.,  2012, \mn@doi [\aap] {10.1051/0004-6361/201118085},
  \href {https://ui.adsabs.harvard.edu/abs/2012A%26A...537A.128R} {537, A128}

\bibitem[\protect\citeauthoryear{{Rein} \& {Spiegel}}{{Rein} \&
  {Spiegel}}{2015}]{Rein2015}
{Rein} H.,  {Spiegel} D.~S.,  2015, \mn@doi [\mnras] {10.1093/mnras/stu2164},
  \href {https://ui.adsabs.harvard.edu/abs/2015MNRAS.446.1424R} {446, 1424}

\bibitem[\protect\citeauthoryear{Rose}{Rose}{2018}]{Rose2018}
Rose B.,  2018, \mn@doi [Journal of Open Source Software]
  {10.21105/joss.00659}, 3, 659

\bibitem[\protect\citeauthoryear{Schwamb et~al.,}{Schwamb
  et~al.}{2013}]{Schwamb2013Kep64Info}
Schwamb M.~E.,  et~al., 2013, \mn@doi [The Astrophysical Journal]
  {10.1088/0004-637x/768/2/127}, 768, 127

\bibitem[\protect\citeauthoryear{{Socia} et~al.,}{{Socia}
  et~al.}{2020}]{Socia2020Kep1661Info}
{Socia} Q.~J.,  et~al., 2020, \mn@doi [\aj] {10.3847/1538-3881/ab665b}, \href
  {https://ui.adsabs.harvard.edu/abs/2020AJ....159...94S} {159, 94}

\bibitem[\protect\citeauthoryear{{Solomon} \& {Head}}{{Solomon} \&
  {Head}}{1991}]{Solomon1991}
{Solomon} S.~C.,  {Head} J.~W.,  1991, \mn@doi [Science]
  {10.1126/science.252.5003.252}, \href
  {https://ui.adsabs.harvard.edu/abs/1991Sci...252..252S} {252, 252}

\bibitem[\protect\citeauthoryear{{Sutherland} \& {Kratter}}{{Sutherland} \&
  {Kratter}}{2019}]{Sutherland2019}
{Sutherland} A.~P.,  {Kratter} K.~M.,  2019, \mn@doi [\mnras]
  {10.1093/mnras/stz1503}, \href
  {https://ui.adsabs.harvard.edu/abs/2019MNRAS.487.3288S} {487, 3288}

\bibitem[\protect\citeauthoryear{{Way}, {Del Genio}, {Kiang}, {Sohl},
  {Grinspoon}, {Aleinov}, {Kelley}  \& {Clune}}{{Way} et~al.}{2016}]{Way2016}
{Way} M.~J.,  {Del Genio} A.~D.,  {Kiang} N.~Y.,  {Sohl} L.~E.,  {Grinspoon}
  D.~H.,  {Aleinov} I.,  {Kelley} M.,   {Clune} T.,  2016, \mn@doi [\grl]
  {10.1002/2016GL069790}, \href
  {https://ui.adsabs.harvard.edu/abs/2016GeoRL..43.8376W} {43, 8376}

\bibitem[\protect\citeauthoryear{Welsh et~al.,}{Welsh
  et~al.}{2012}]{Welsh2012Kep35Info}
Welsh W.~F.,  et~al., 2012, Nature, 481, 475 EP

\bibitem[\protect\citeauthoryear{{Welsh} et~al.,}{{Welsh}
  et~al.}{2015}]{Welsh2014Kep453Info}
{Welsh} W.~F.,  et~al., 2015, \mn@doi [\apj] {10.1088/0004-637X/809/1/26},
  \href {https://ui.adsabs.harvard.edu/abs/2015ApJ...809...26W} {809, 26}

\bibitem[\protect\citeauthoryear{{Williams} \& {Pollard}}{{Williams} \&
  {Pollard}}{2002}]{Williams2002}
{Williams} D.~M.,  {Pollard} D.,  2002, \mn@doi [International Journal of
  Astrobiology] {10.1017/S1473550402001064}, \href
  {https://ui.adsabs.harvard.edu/abs/2002IJAsB...1...61W} {1, 61}

\makeatother
\end{thebibliography}

\begin{figure}
\centering
\includegraphics[width=0.9\linewidth]{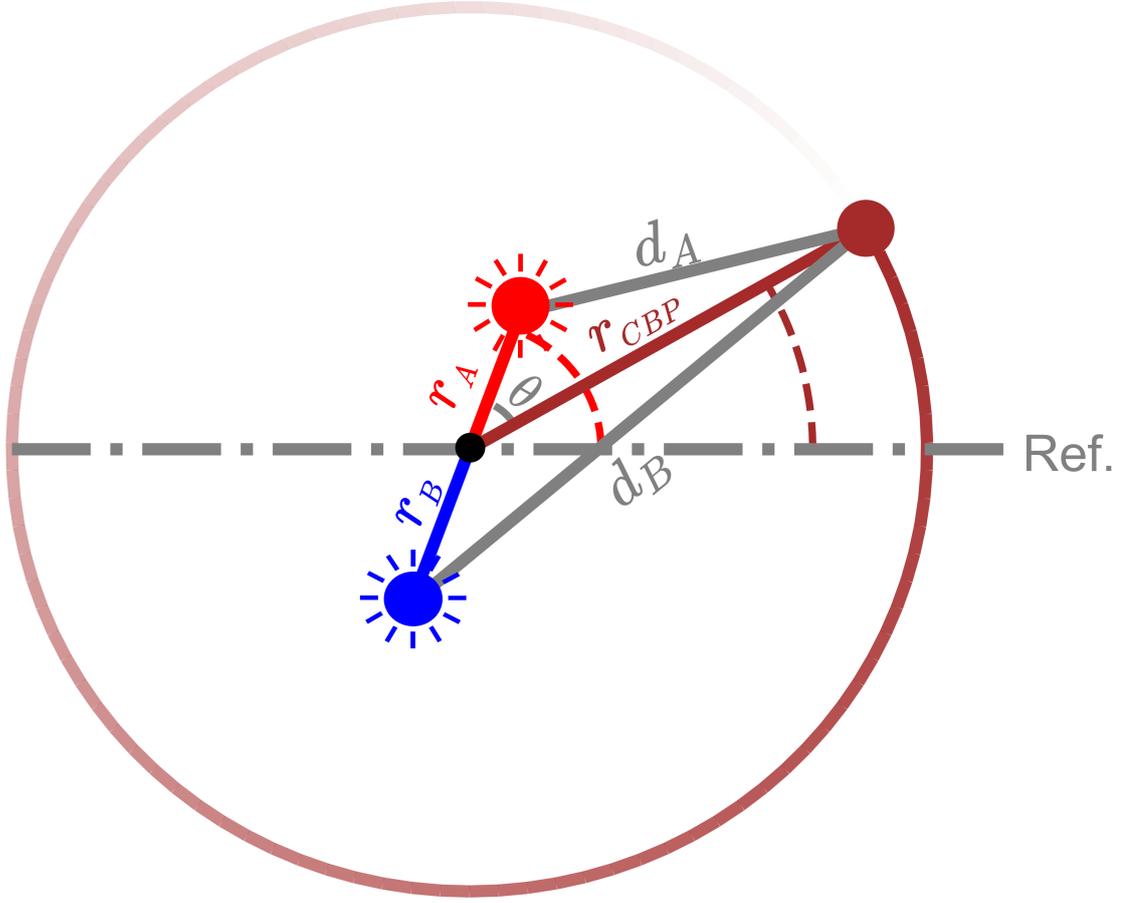}
\caption{Visualization of the CBP-binary configuration, with the the primary star (red), secondary star (blue), and CBP (brown) color-coded.  The distances from center of mass to the primary star, secondary star, and CBP are $r_A$, $r_B$, $r_{CBP}$, respectively. The relative distance between the CBP to the primary and secondary are $d_A$ and $d_B$, respectively. The angular displacement of the primary star ($\omega_{bin} + f_{bin}$) and the CBP ($\omega_{CBP} + f_{CBP}$) from the reference direction are shown by the red dashed and the brown dashed curves, respectively. The angle between the position vectors of the primary star and the CBP is denoted by $\theta$.}
\label{fig:CBPConfig}
\end{figure}

\begin{figure}
\centering
\includegraphics[width=0.9\linewidth]{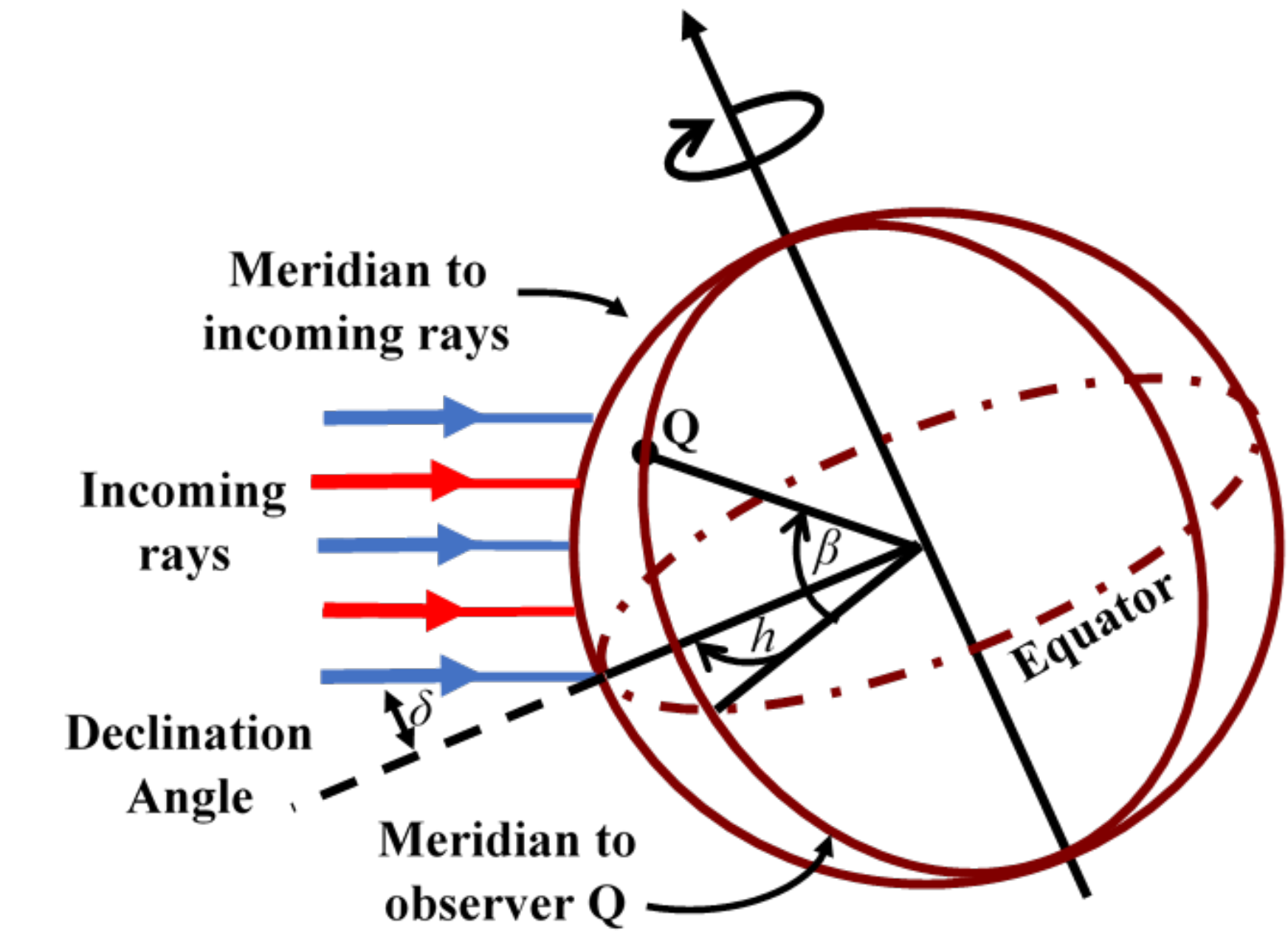}
\caption{Visualization of the received flux from the perspective of an observer {Q} on the surface of a CBP, with latitude ($\beta$), hour angle ($h$), and declination ($\delta$). }
\label{fig:CBPAngles}
\end{figure}

\begin{figure}
    \centering
   \includegraphics[width=0.7\linewidth]{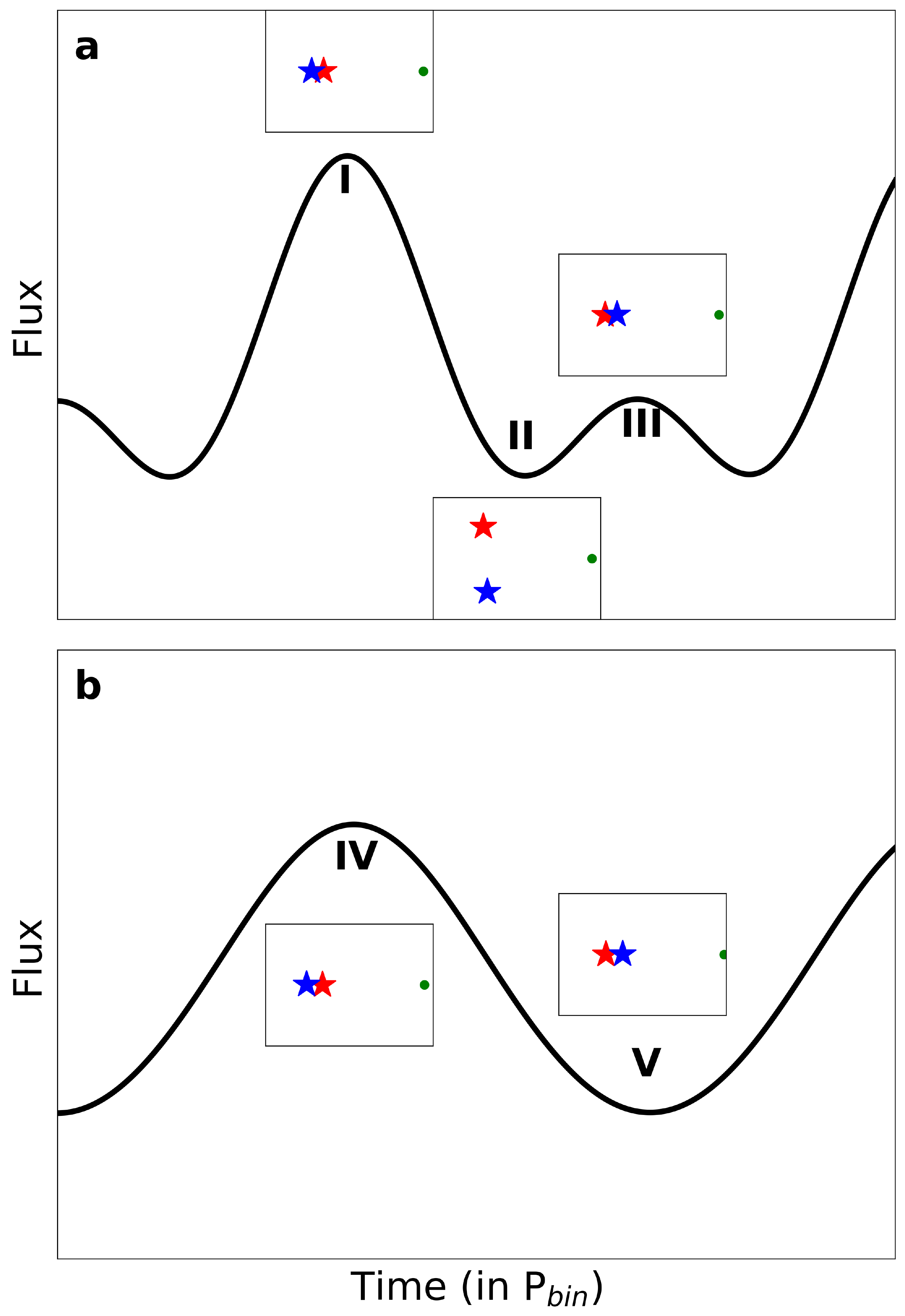}

\caption{These are the {top-down views of} configurations of a {Case 1} CBP-Binary system that produce a maximum (I), minimum (II), and a secondary maximum (III) flux on the CBP's surface (top), and a {Case 2} CBP-Binary system that produce maximum (IV) and minimum (V) flux on the CBP's surface (bottom). The red star is the primary and the blue star is the secondary.}
\label{fig:FluxMaxMinCBP} 
\end{figure}

\begin{figure}
    \centering
       \includegraphics[width=\linewidth]{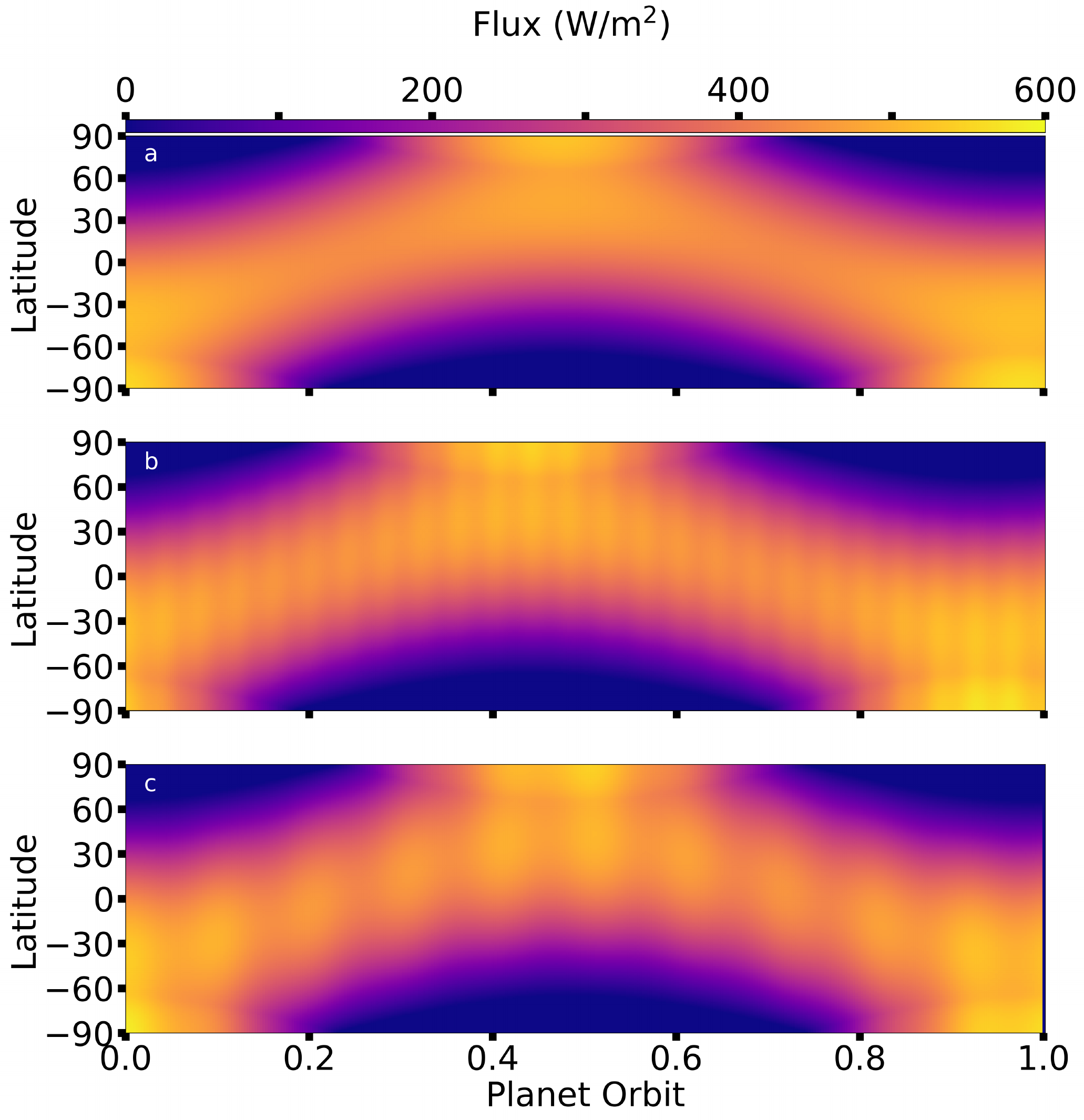}
      
\caption{Flux as a function of latitude on an Earth-like planet receiving an Earth flux orbiting (a) a Sun-like star, (b) a \textit{Case 1} CBP configuration, and (c) a \textit{Case 2} CBP configuration.  The inner binary for the CBP configurations begin on a circular orbit with a binary semimajor axis $a_{bin} =$ 0.45 au.}
\label{fig:Earth-CBPDiff}
\end{figure}

\begin{figure}
   \centering
       \includegraphics[width=\linewidth]{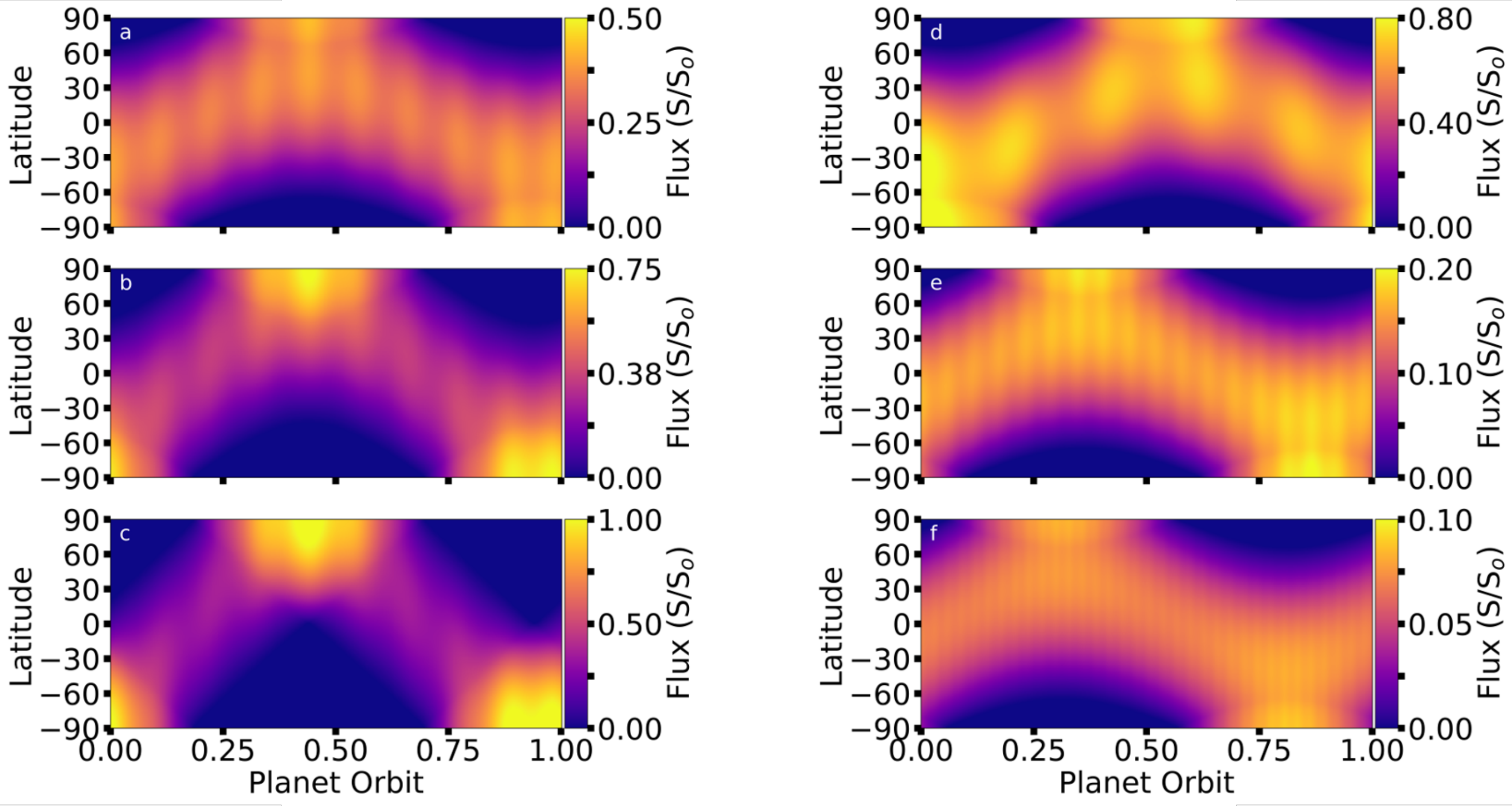}

\caption{Flux changes as a function of obliquity for: (a) 23.5$^\circ$, (b) 45$^\circ$, and (c) 90$^\circ$ and as a function of planetary semimajor axis $a_p$ for: (d) $2a_{crit}$, (e) $3a_{crit}$, and (f) $4a_{crit}$. The orbit of the CBP and its Sun-like binary host stars begin on circular orbits, where the binary separation $a_{bin}$ is 0.45 au.}
\label{fig:FluxCBPSemimObl} 
\end{figure}

\begin{figure}
\includegraphics[width=\linewidth]{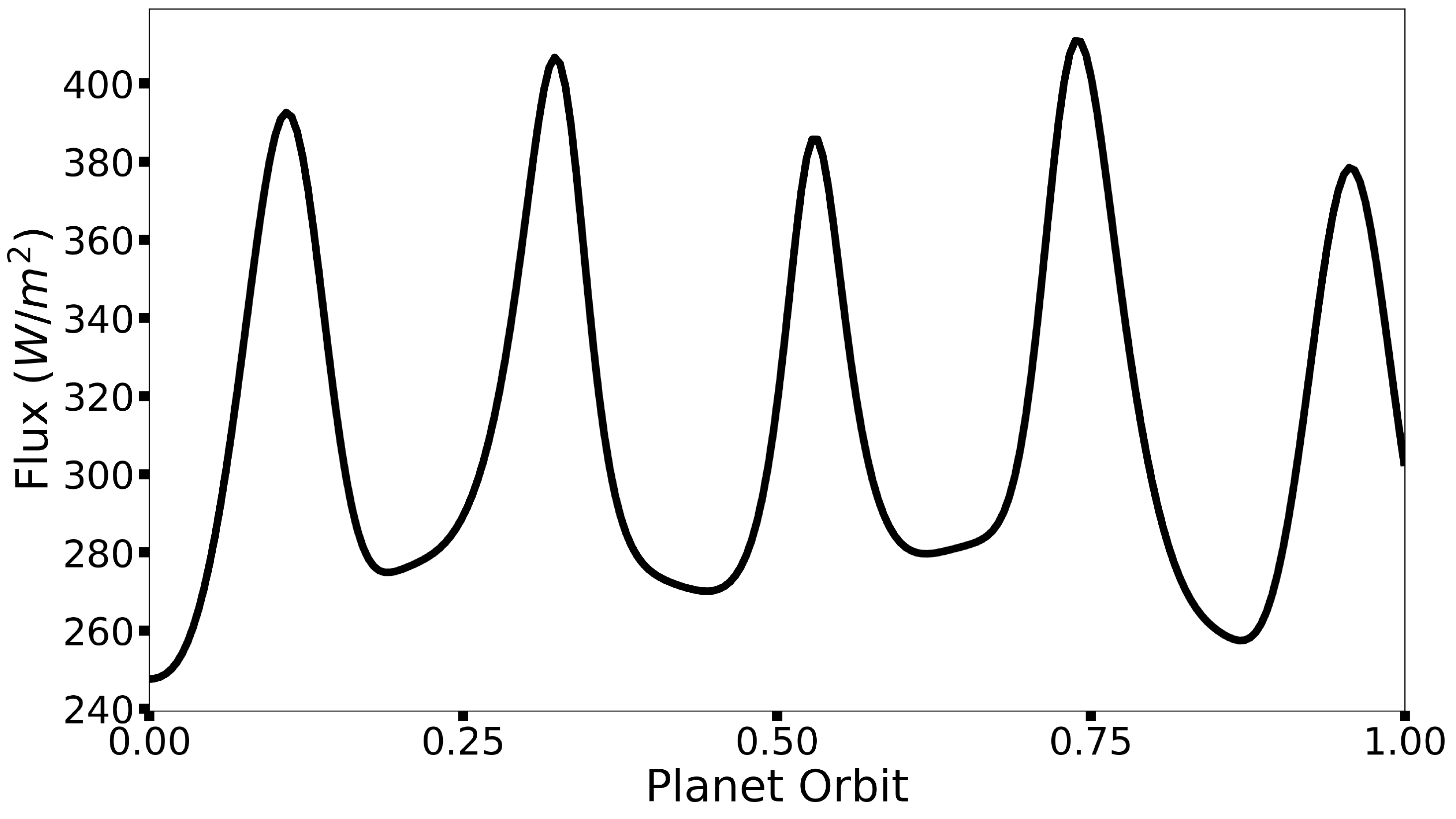}
\caption{The top of the atmosphere flux over the equator of Kepler-16b as a function of its orbit around the Kepler-16 G-M binary (a \textit{Case 2} configuration).
\label{fig:FluxSA250}}
\end{figure}

\begin{figure}
\includegraphics[width=\linewidth]{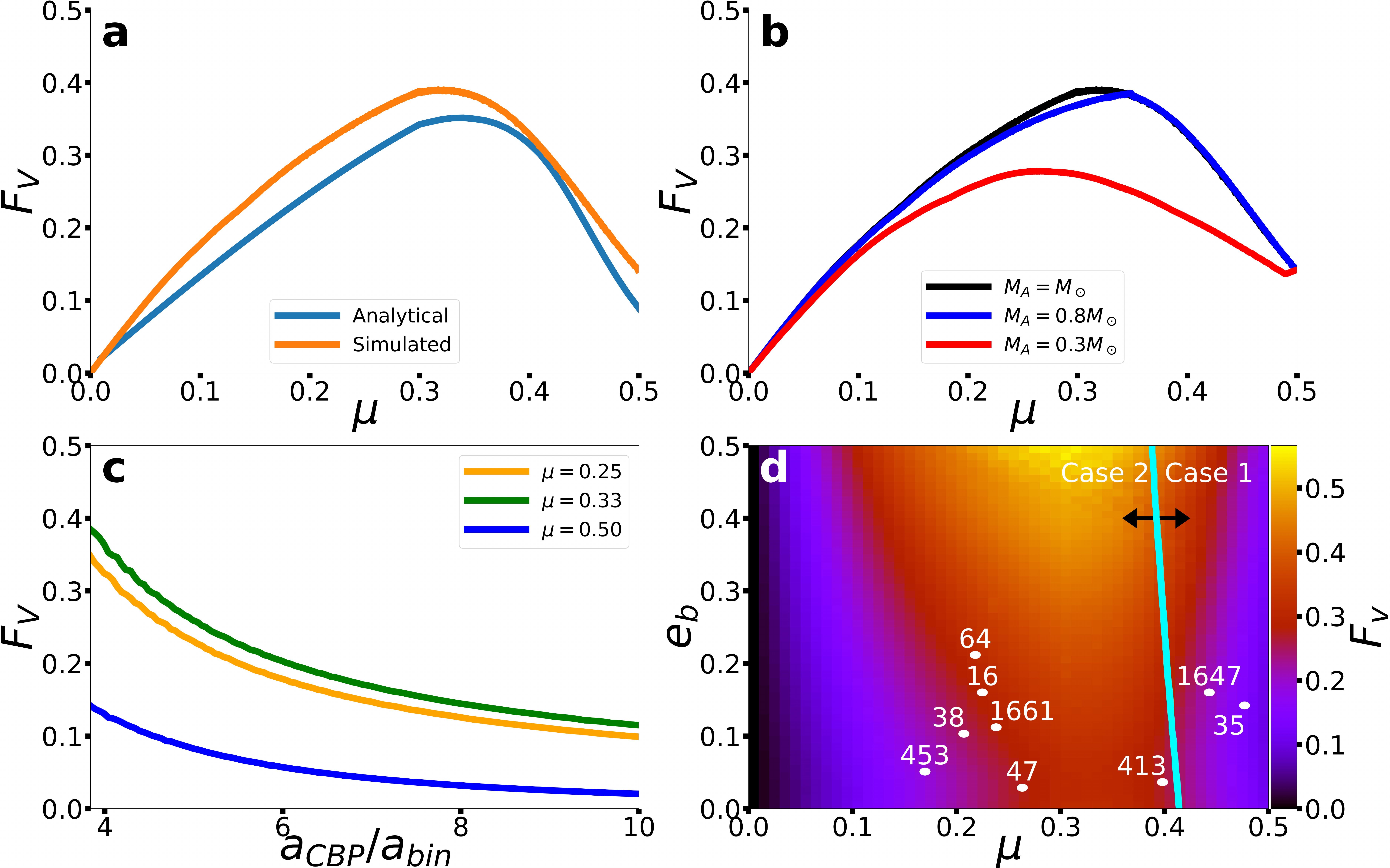}
\caption{In panel (a), we show analytical and simulated calculations of flux variation over dynamical mass ratio $\mu$, where the primary is Sun-like. Panel (b) shows simulated values of $F_V$ using a range of primary host stars (0.3 M$_\odot$, 0.8 M$_\odot$, and 1 M$_\odot$). Panel (c) illustrates how $F_V$ decreases with respect to $a_{CBP}/a_{bin}$ from $3.84-10$.  For panel (d), we show flux variations on a CBP over a range of binary eccentricity $e_{bin}$ and mass ratios $\mu$.  Panels (a), (b) and (c) use a stellar orbit with a typical semimajor axis ($a_{bin} = 0.15$) and eccentricity ($e_{bin} = 0.13$) characteristic of known CBP host binaries.  The CBPs in panels (a), (b), and (d) begin on a circular orbit beyond the stability limit at 0.576 AU.}
\label{fig:FluxVarMu}
\end{figure}

\begin{figure}
    \centering
       \includegraphics[width=\linewidth]{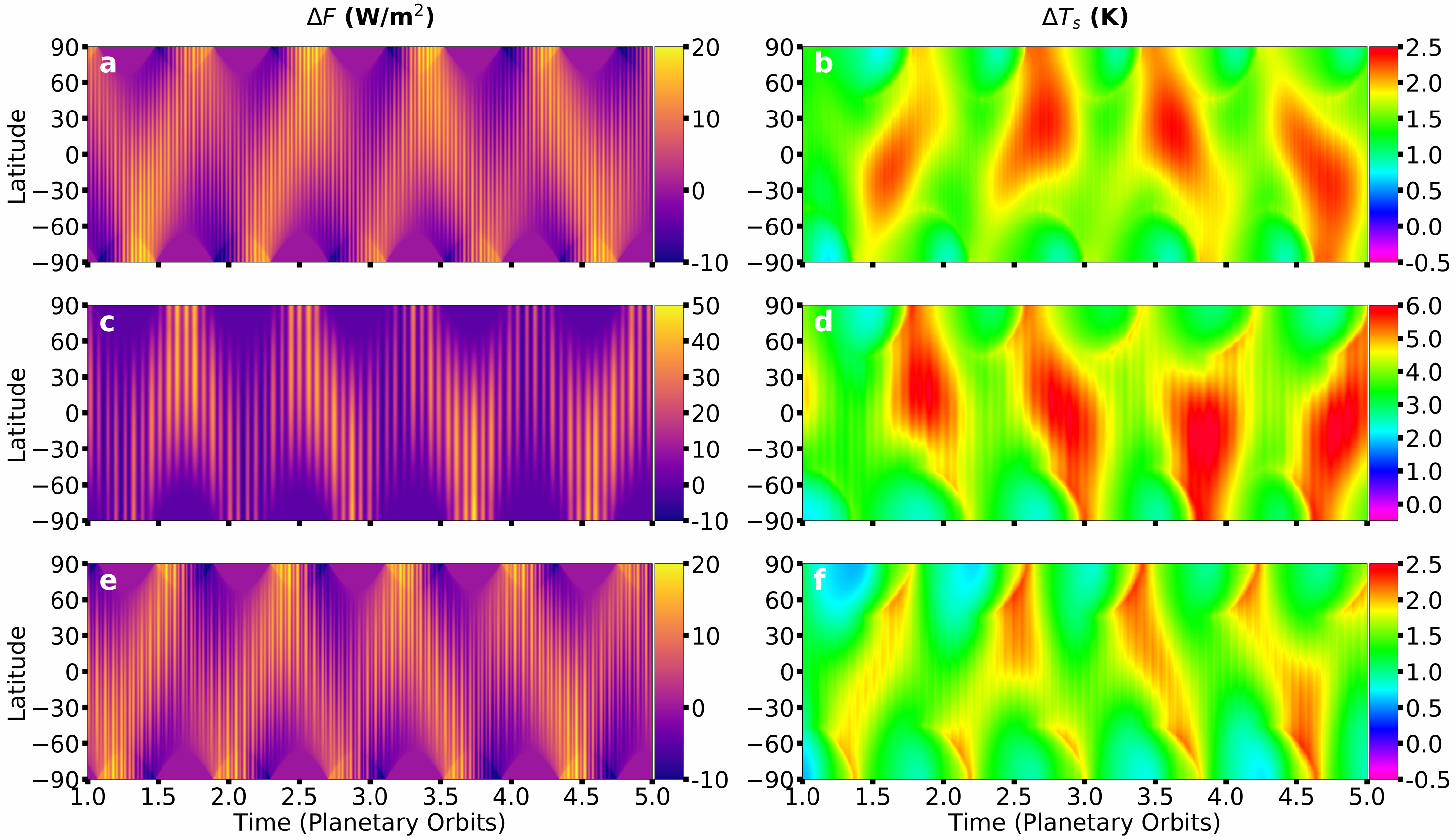}
      
\caption{The differential flux ($\Delta F = F_{CBP} - F_{ESS}$, left column) and surface temperature ($\Delta T_s = T_{CBP} - T_{ESS}$, right column) between a \textit{Case 1} CBP (equal mass stars, total luminosity of $L_\odot$) and the equivalent single-star (ESS) system over 4 planetary orbits. Panels (a) and (b) illustrate the differential values using binary parameters that are typical of the known CBP hosts. Panels (c) and (d) illustrate the changes in differential flux and surface temperature when the binary period is doubled. Panels (e) and (f) show the changes in differential flux and surface temperature when binary eccentricity $e_{bin}$ is doubled instead.  The CBP in each scenario starts on an orbit receiving one Earth flux, with an Earth-like obliquity (23.5$^\circ$) and on a circular orbit (see Table \ref{table:Case_EBM}). }
\label{fig:FluxTempDeltaCase1}
\end{figure}

\begin{figure}
    \centering
       \includegraphics[width=\linewidth]{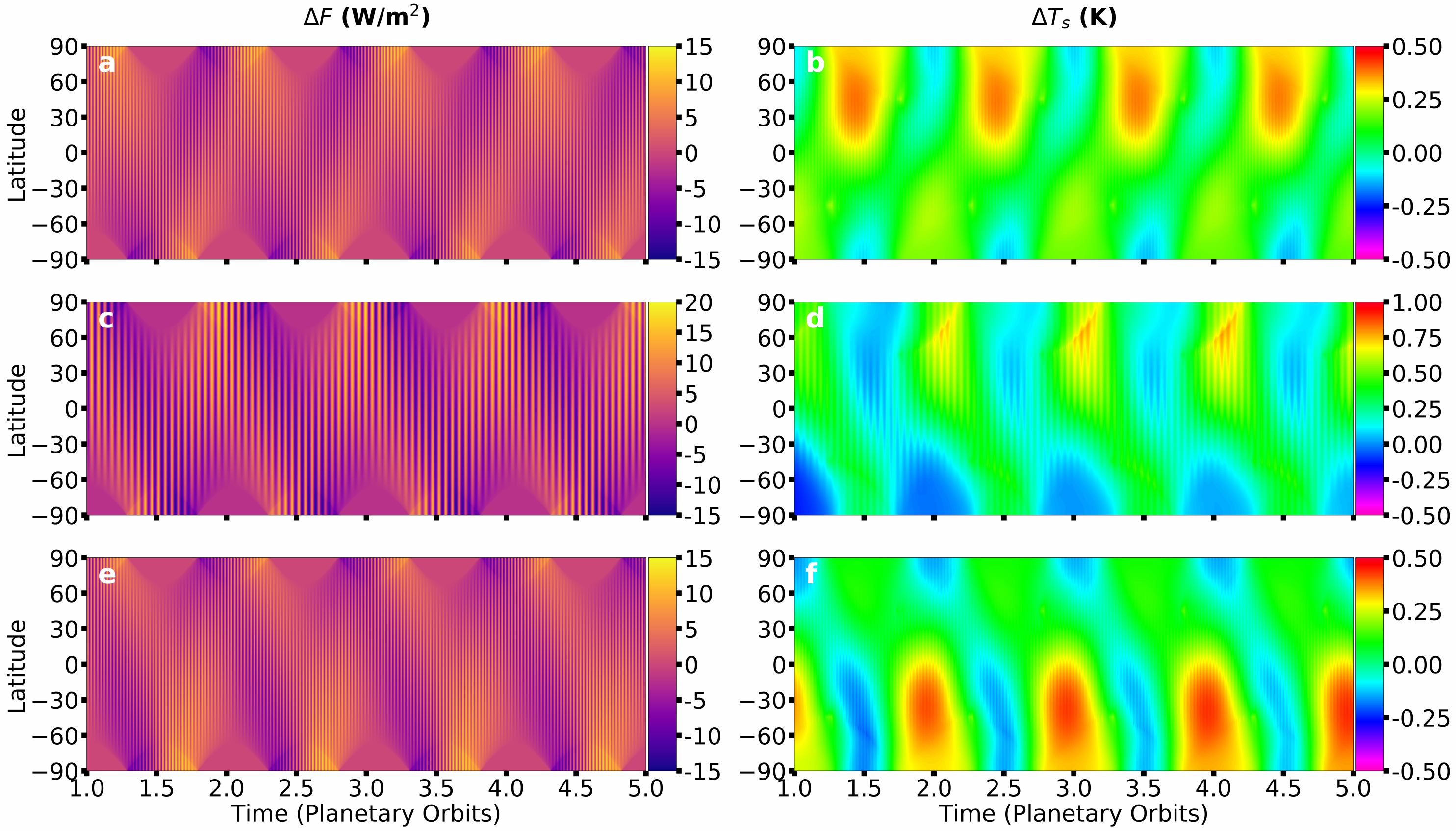}
      
\caption{Similar to Figure \ref{fig:FluxTempDeltaCase1}, but for a \textit{Case 2} CBP configuration (see Table \ref{table:Case_EBM}).}
\label{fig:FluxTempDeltaCase2}
\end{figure}

\begin{figure}
\centering
\includegraphics[width=0.8\linewidth]{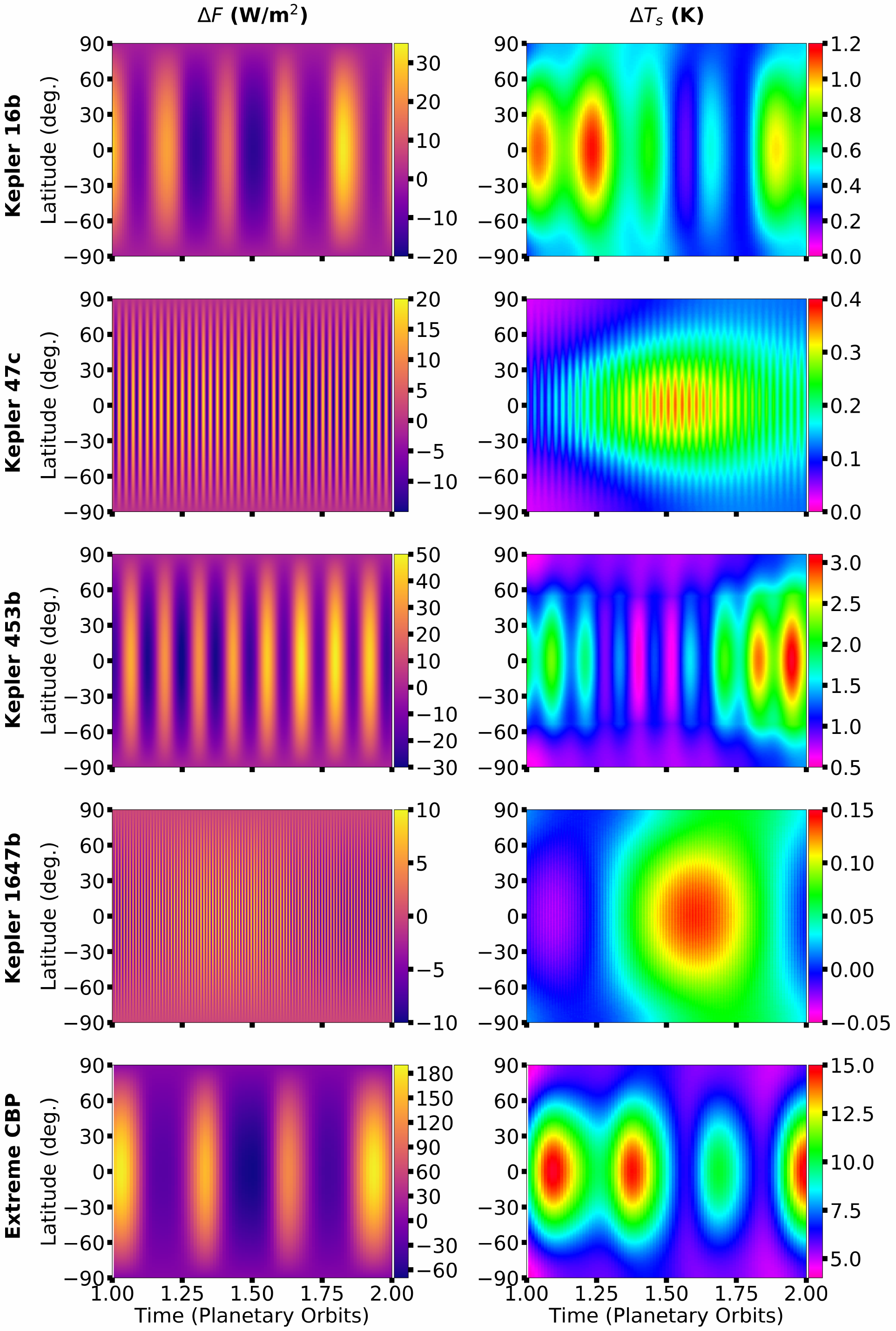}
\caption{Differential flux ($\Delta F = F_{CBP} - F_{ESS}$) and surface temperature ($\Delta T_s = T_{CBP}-T_{ESS}$) for the four habitable zone CBPs with their equivalent single-star (ESS) counterparts over an orbit. An additional case is provided for an extreme CBP, where its semimajor axis is near the stability limit and the inner edge of its HZ (see Table \ref{table:Plot8CBPValues}).}  
\label{fig:Kep_CBP_time}
\end{figure}

\begin{figure}
\centering
\includegraphics[width=\linewidth]{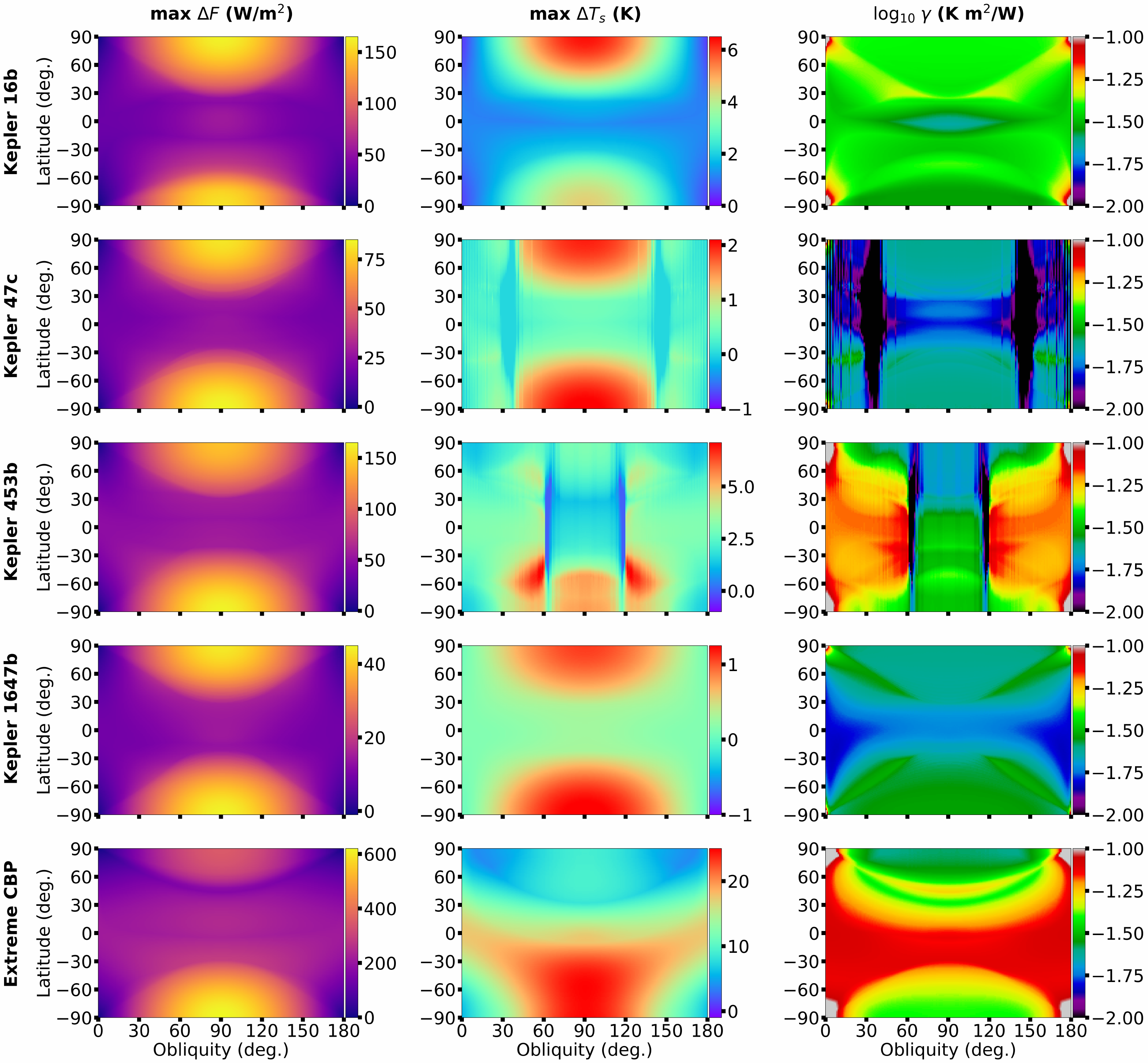}
\caption{The maximum differential flux ($\Delta F = F_{CBP} - F_{ESS}$) and surface temperature ($\Delta T_s = T_{CBP}-T_{ESS}$) contrasts the four habitable zone CBPs and one hypothetical case (Extreme CBP) with their equivalent single-star (ESS) counterparts as a function of the planetary obliquity.  The third column illustrates ratio $\gamma$, which equals the (max $\Delta F$)/(max $\Delta T_s$), on a base-10 logarithmic scale.  The physical meaning of $\gamma$ is a measure of how much of a surface temperature change ($\Delta T_s$) results from an additional flux ($\Delta F$) from the CBP host stars.}
\label{fig:Kep_CBP_max}
\end{figure}

\begin{figure}
\centering
\includegraphics[width=0.8\linewidth]{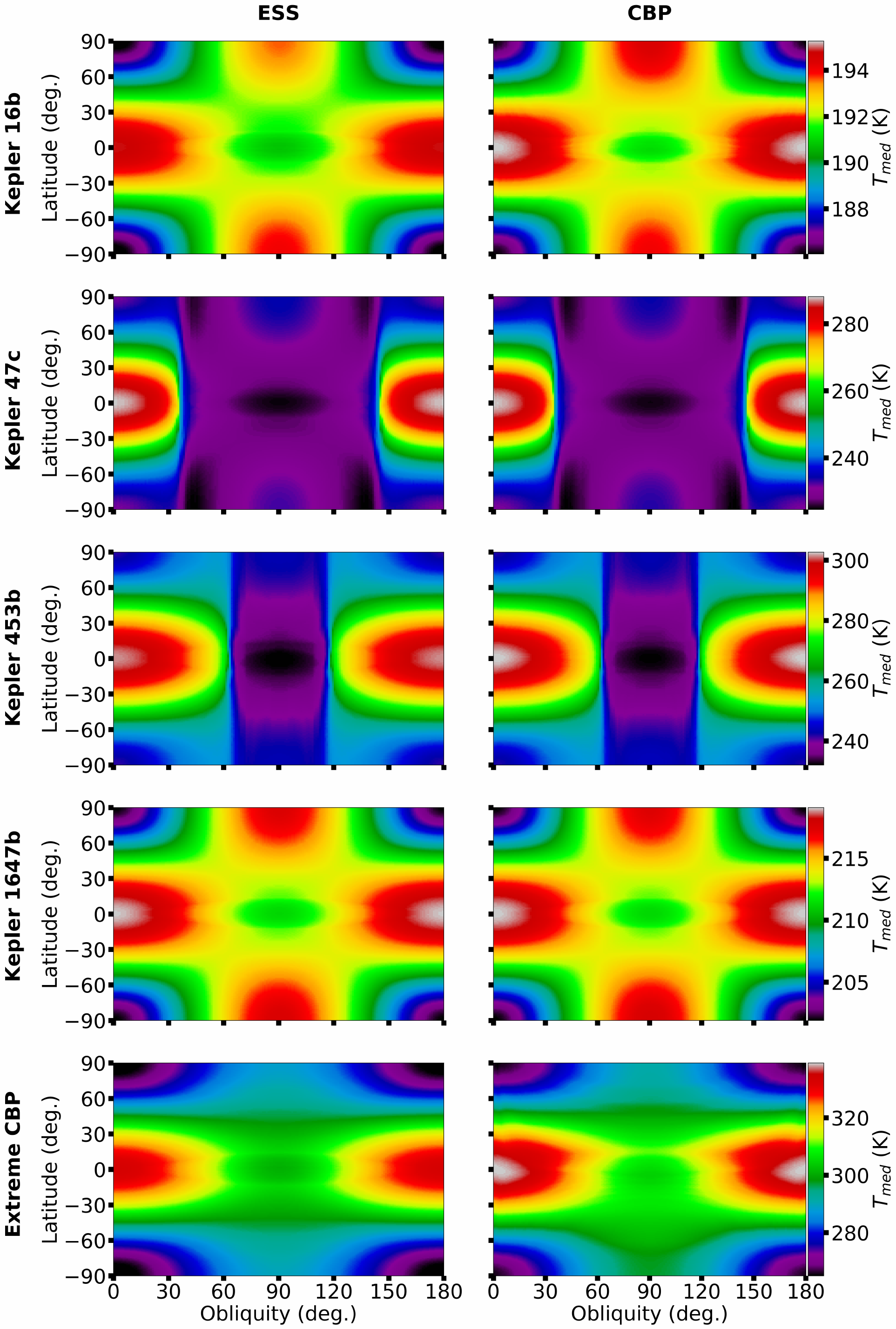}
\caption{The median surface temperature ($T_{med}$) experienced over an orbit by the habitable zone CBPs and one hypothetical case (Extreme CBP) compared to their equivalent single-star (ESS) counterparts as a function of the planetary obliquity.  Each row uses the same color scale located on the right.  Note that the CBPs are typically warmer than the ESS configuration and the gyration from the binary reduces the efficacy of the ice-albedo feedback at high ($\sim$90$^\circ$) obliquities.}
\label{fig:Kep_CBP_med}
\end{figure}

\begin{figure}
\includegraphics[width=\linewidth]{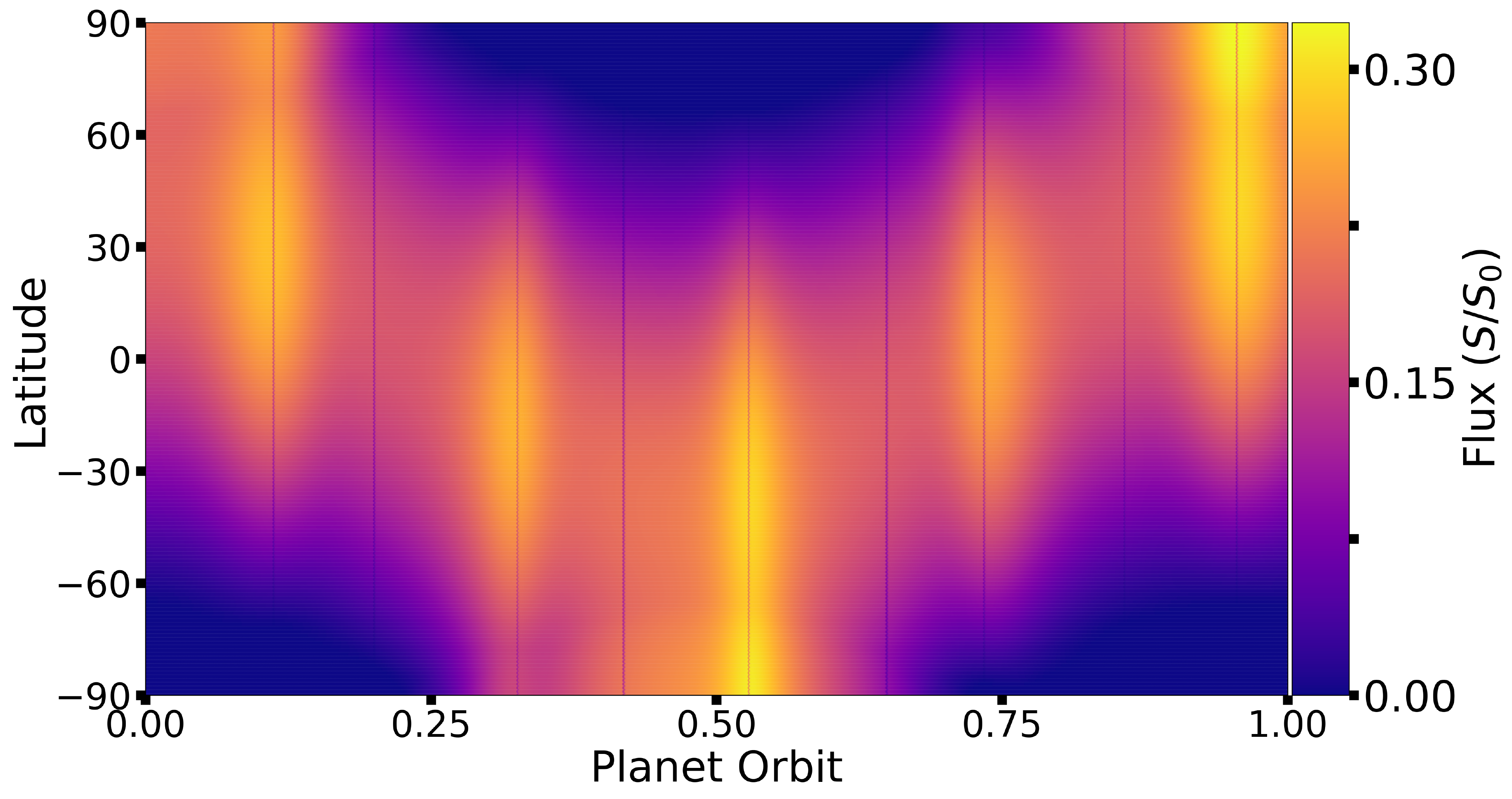}
\caption{This is the top of atmosphere flux received at all latitudes of Kepler-16b as a function of its orbit, accounting for eclipses. The time resolution plotted is 0.1 Earth day.}
\label{fig:FluxLatEclipses}
\end{figure}

\newpage
{\appendix
\section{Deriving the Border Between Case 1 and 2}
\label{BorderDeriveAppendix}
The border between \textit{Case 1} and \textit{Case 2} can be delineated by using orbital elements that parameterize the system relative to the center of mass.  Figure \ref{fig:CBPConfig} illustrates this process through the distance from the center of mass to star A $r_A$ (red), star B $r_B$ (blue), and the CBP $r_{CBP}$ (maroon).  In addition to these parameters, the distance from the CBP to star A $d_A$ and star B $d_B$ are denoted in gray.  We represent the orientation of the CBP orbit ($\omega_{CBP} + f_{CBP}$; maroon) and the binary orbit ($\omega_{bin}+f_{bin}$; red) through color-coded dashed curves.  Finally, we identify the angle between the position vectors of the primary star A and the CBP by the symbol $\theta$ (gray).  The border between \textit{Case 1} and \textit{Case 2} can be identified through the second derivative of the weighted flux (i.e., $\frac{\partial^2 F}{\partial \theta^2} = 0$) when the secondary star B eclipses the primary star A ($\theta = \pi$), where the weighted flux is:

\begin{equation} \label{eqn:tot_flux}
    F = \frac{L_A}{4\pi d_A^2} + \frac{L_B}{4\pi d_B^2}.
\end{equation}
The stellar luminosities $L_A$ and $L_B$ should be adjusted for their spectral weight factor \citep{Haghighipour2013} at this point or by multiplying our final result by $W^x_{Sec}/W^x_{Pr}$.  The distances to each of the stars ($d_A$ and $d_B$) can be determined using the Law of Cosines relative to the center of mass and we demonstrate this calculation for $d_A$:

\begin{equation} \label{eqn:cosine_law}
    d_A^2 = r_{CBP}^2 + r_A^2 - 2 r_{CBP} r_A \cos \theta,
\end{equation}

where the distances $r_{CBP}$ and $r_A$ relative to the center of mass are found using the radial equation from the two body problem \citep{Murray1999}.  Using these analytic expressions is acceptable when the CBP is far enough from its inner stability limit \citep{Quarles2018} so that changes to the CBP's semimajor axis and eccentricity are small over a planetary orbit.  In addition, the radial equations require adjustment so that the cosine term includes the angle $\theta$.  To simplify the problem, we set $\omega_{CBP} = \omega_{bin} = 0$, which gives a measure of the relative phase $\theta$ between the orbits ($\theta = f_{bin} - f_p$).  As a result, the radial equations for $r_{CBP}$ and $r_A$ are

\begin{equation}
r_{CBP}=\frac{a_{CBP}(1 - e_{CBP}^2)}{1 + e_{CBP} \cos(f_{bin}-\theta)}
\end{equation}

and 

\begin{equation}
r_A=\frac{\mu a_{bin}(1 - e_{bin}^2)}{1 + e_{bin} \cos(\theta+f_p)},
\end{equation}
which includes the dynamical mass ratio $\mu$ for the binary orbit.  The cosine rule (equation \ref{eqn:cosine_law}) calculates $d_A^2$, where $1/d_A^2$ is required in equation \ref{eqn:tot_flux}.  At this point, it is likely more practical and accurate to calculate $d_A^2$ numerically using \texttt{REBOUND}.  However, we can employ the method of expansion of a $1/r$ potential into Legendre polynomials and discard the appropriate terms, where such an analytical approximation could be more useful when the number of trial cases is large.  Thus, we re-arrange equation \ref{eqn:cosine_law} into the following form:

\begin{equation} \label{eqn:Legendre}
\frac{1}{d_A} = \frac{1}{r_{CBP}}\left[1+\left(\frac{r_A}{r_{CBP}} \right)^2 - 2\frac{r_A}{r_{CBP}}\cos \theta \right]^{-0.5} = \frac{1}{r_{CBP}}\sum_{n=0}^{\infty}\left(\frac{r_A}{r_{CBP}}\right)^n \mathcal{P}_n(\cos \theta),
\end{equation}

where $\mathcal{P}_n$ is the $n$-th Legendre polynomial.  Assuming that the ratio $r_A/r_{CBP} << 1$ (a requirement for orbital stability), we set $\chi= r_A/r_{CBP}$ and expand equation \ref{eqn:Legendre} to second order in $\chi$ to get:

\begin{equation} \label{eqn:Leg_exp}
\frac{1}{d_A} \approx \frac{1}{r_{CBP}}\left[1+\chi\cos \theta+\frac{1}{2}\chi^2(3\cos^2\theta-1) + \mathcal{O}\left(\chi^3\right)\right],
\end{equation}
where $1/d_A^2$ can be found by squaring equation \ref{eqn:Leg_exp} and removing the higher order terms.  The final approximation for $1/d_A^2$ is:
\begin{equation} \label{eqn:Leg_exp_sqr}
\frac{1}{d_A^2} \approx \frac{1}{r_{CBP}^2}\left[1+2\chi\cos\theta+\chi^2(4\cos^2\theta-1) \right],
\end{equation}

Most CBPs of interest (i.e., within their HZ) are expected to have low eccentricity \citep[][their Fig. 8]{Quarles2018} and thus, we set $e_{CBP}$ to zero.  Moreover, we can choose the reference line such that $f_{CBP}$ is also zero, which simplifies equation \ref{eqn:Leg_exp_sqr} greatly.  The expansion to find $d_B$ follows the same procedure (equations \ref{eqn:cosine_law}--\ref{eqn:Leg_exp_sqr}) through the substitutions: $f_{bin} \rightarrow f_{bin} + \pi$ and $\mu \rightarrow (1-\mu)$.  Evaluating $\frac{\partial^2 F}{\partial \theta^2} = 0$ at $\theta = \pi$, we find the critical luminosity ratio between \textit{Case 1} and \textit{Case 2} as:

\begin{equation}
\frac{L_B}{L_A} = \frac{\mu [1 + \alpha (e_{bin} - 4) (e_{bin} + 1) \mu]}{[1 - \alpha (e_{bin} - 4) (e_{bin} + 1) (1 - \mu)] (1 - \mu)},
\end{equation}
where $\alpha$ denotes the semimajor axis ratio $a_{bin}/a_{CBP}$.}

\clearpage
\newpage

\bsp	
\label{lastpage}
\end{document}